# A 3D view of the Hyades stellar and sub-stellar population

N. Lodieu[1,2,3], R. L. Smart[3], A. Pérez-Garrido[4], and R. Silvotti[3]

[1] Instituto de Astrofísica de Canarias (IAC), Calle Vía Láctea s/n, E-38200 La Laguna, Tenerife, Spain
e-mail: nlodieu@iac.es
[2] Departamento de Astrofísica, Universidad de La Laguna (ULL), E-38206 La Laguna, Tenerife, Spain
[3] Istituto Nazionale di Astrofisica, Osservatorio Astrofisico di Torino, Strada Osservatorio 20, I-10025 Pino Torinese, Italy
[4] Dpto. Física Aplicada, Universidad Politécnica de Cartagena, E-30202 Cartagena, Murcia, Spain

## ABSTRACT

*Aims.* Our scientific goal is to provide a 3D map of the nearest open cluster to the Sun, the Hyades, combining the recent release of *Gaia* astrometric data, ground-based parallaxes of sub-stellar member candidates and photometric data from surveys which cover large areas of the cluster.
*Methods.* We combined the second *Gaia* release with ground-based *H*-band parallaxes obtained with the infrared camera on the 2-m robotic Liverpool telescope to astrometrically identify stellar and sub-stellar members of the Hyades, the nearest open cluster to the Sun.
*Results.* We find 1764 objects within 70° radius from the cluster center from the *Gaia* second data release, whose kinematic properties are consistent with the Hyades. We limit our study to 30 pc from the cluster center (47.03±0.20 pc) where we identify 710 candidate members, including 85 and 385 in the core and tidal radius, respectively. We determine proper motions and parallaxes of eight candidate brown dwarf members and confirm their membership. Using the 3D positions and a model-based mass-luminosity relation we derive a luminosity and mass function in the 0.04 to 2.5 $M_\odot$ range. We confirm evidence for mass segregation in the Hyades and find a dearth of brown dwarfs in the core of the cluster. From the white dwarf members we estimate an age of $640^{+67}_{-49}$ Myr.
*Conclusions.* We identify a list of members in the Hyades cluster from the most massive stars down to the brown dwarfs. We produce for the first time a 3D map of the Hyades cluster in the stellar and sub-stellar regimes and make available the list of candidate members.

## 1. Introduction

The large majority of stars are born in groups, clusters, and/or associations rather than in isolation (Lada & Lada 2003). Investigating the dynamics of clusters is key to understanding their formation and subsequent evolution. Up to now, ground-based surveys mainly provided proper motion information as well as partial radial velocity measurements offering a two-dimensional map of the sky and, in particular for this work, of nearby open clusters. Using accurate astrometric observations of the Hipparcos satellite ((High precision parallax collecting satellite; Perryman et al. 1997), a catalogue of about 120,000 stars brighter than visual magnitude $V = 12.4$ mag with a completeness limit of $V \sim 8$ mag was generated. Later, the Tycho-2 catalogue was created containing 2.5 million stars down to $V = 11.5$ and 99% complete to $V \sim 10.5$ mag over the full sky (Høg et al. 2000). The advent of *Gaia* (Gaia Collaboration et al. 2016) is opening an unprecedented window with accurate proper motions and parallaxes for more than one billion stars down to $G \sim 20.7$ mag, allowing a spatial investigation of our Galaxy at large radii never looked at before at that level of precision. In particular, *Gaia* provides accurate 3D space motions for the nearest and youngest open clusters to the Sun, including the Hyades.

The Hyades cluster (M25, Collinder 50, Caldwell 41) is the closest stellar cluster to the Sun. Using the Hipparcos catalogue Perryman et al. (1998) derived a mean distance of 46.34±0.27 pc and a proper motion in the 74–140 mas/yr range. The core radius of the cluster is approximately 2.5–3.0 pc while its tidal radius is about 10 pc (Perryman et al. 1998; Röser et al. 2011). The reddening towards the cluster is negligible (E(*B*−*V*) ≤0.001 mag; Taylor 2006a).

The age of the cluster has been estimated with different methods, yielding a canonical age of 650±100 Myr. Comparison of the observed cluster sequence with model isochrones which include convective overshooting suggests 625±50 Myr while evolutionary models (Maeder & Mermilliod 1981; Mazzei & Pigatto 1988; Mermilliod 1981) with enhanced convective overshooting give much older ages up to 1.2 Gyr (Mazzei & Pigatto 1988) that cannot be discarded (Eggen 1998; Tremblay et al. 2012). The ages determined from the cooling age of white dwarf members is 648±45 Myr (De Gennaro et al. 2009) while stellar binaries suggest ~650 Myr (Lebreton et al. 2001). The role of rotation at such ages leads to an age of 750±100 Myr (Brandt & Huang 2015). Recently, the method using the lithium depletion boundary at the stellar/sub-stellar limit prompted an age of 650±70 Myr consistent with the canonical age of the cluster (Lodieu et al. 2018; Martín et al. 2018).

The metallicity has been subject to debate with discrepant estimates suggesting a mean metallicity close to solar (Fe/H = 0.05±0.05; Gebran et al. 2010) or slightly super-solar between 0.127±0.022 dex and 0.14±0.10 dex (Boesgaard & Friel 1990; Cayrel de Strobel et al. 1997; Grenon 2000).

The surveys targeting the Hyades can be divided up into two main groups: the large-scale studies looking for a complete census of the cluster members over very large areas of the sky (Gizis et al. 1999; Goldman et al. 2013; Hanson 1975; Hogan et al. 2008; Reid 1992; Röser et al. 2011) or deeper surveys on small(er) patches in the cluster center to identify new members (Bouvier et al. 2008; Dobbie et al. 2002; Hanson 1975; Leggett et al. 1994; Melnikov & Eislöffel 2018; Reid & Gizis 1997; Reid et al. 1999; Reid 1993; Stauffer et al. 1995, 1994). Subsequent spectroscopic follow-up has been conducted for many sources to confirm their membership via spectral typing, radial velocity and/or lithium content (Bryja et al. 1994; Leggett & Hawkins 1989; Mermilliod et al. 2009; Reid & Hawley 1999; Reid & Mahoney 2000; Soderblom et al. 1995; Stauffer et al. 1995, 1994; Tabernero et al. 2012; White et al. 2007). The current census down to approximately 0.1 $M_\odot$ is summarised in Röser et al. (2011) using the Positions and Proper Motion Extra Large catalogue (PPMXL; Röser et al. 2010) and the Panoramic Survey Telescope and Rapid Response System first data release (Pan-STARRS DR1; Chambers et al. 2016; Goldman et al. 2013; Kaiser et al. 2002) complemented by the L and T dwarf member candidates analysed by Hogan et al. (2008) and Bouvier et al. (2008), respectively. The coolest members have been confirmed spectroscopically with masses below the hydrogen-burning limit (Bouvier et al. 2008; Casewell et al. 2014; Lodieu et al. 2014a, 2018; Martín et al. 2018).

In this manuscript, we present an astrometric selection of Hyades cluster member candidates from the second *Gaia* release (DR2; Gaia Collaboration et al. 2018b) yielding a revised census and a 3D map with positions in the sky. In Sect. 2 we present the input catalogue used for our study of the Hyades cluster and compile a list of previously-known members in Sect. 3. In Sect. 4 we describe complementary ground-based parallaxes from a dedicated program carried out with the Liverpool telescope for the coolest member candidates of the cluster. In Sect. 5 we identify member candidates in the Hyades from *Gaia* DR2 and compare with previous studies to address the completeness and contamination of our sample. We dedicate Sect. 6 to the analysis of white dwarf members and their implication on the age of the cluster that we compare with other estimates. In Sect. 7 we derive the luminosity and mass functions in the stellar and substellar regimes. In Sect. 8 we discuss the spatial distribution of the highest probability member candidates and present the first 3D map of the Hyades from the most massive members down to the sub-stellar regime.

## 2. The *Gaia* DR2 sample

We made use of the *Gaia* DR2 data (Gaia Collaboration et al. 2018b, 2016) released world-wide on April 25$^{th}$ 2018. Our objective was to start as inclusive as possible and to be more selective later in the process. Initially we started with all *Gaia* DR2 objects that were within 70° of the nominal cluster center at $\alpha = 67.0°$, $\delta = +16.0°$ and had a parallax greater the 10 mas, i.e. within 100 pc, which results in 126,144 objects (Fig. 1). We believe that the Hyades cluster has a on sky dimension of less than 30 pc (Röser et al. 2011). By extending the search from 30 to 70°, our analysis shall include all possible members with measured parallaxes but we will restrain our analysis to a volume of 30 pc from the cluster center.

We cross-matched this catalogue with a serie of well-known large-scale surveys to provide photometry over a wide wavelength range, keeping all *Gaia* DR2 sources without any counterpart in those surveys either due to coverage or bright-ness/faintness reasons. The best match was obtained with a maximum distance of 10″. We did not use the official *Gaia* DR2 cross-match tables but our own script with `stilts` (Taylor 2006b) because not all public large-scale catalogues are available with the online archive. All the unresolved sources in the large-scale surveys that were resolved in *Gaia* were matched to the closest *Gaia* source which is normally the brighter of the two possible matches. All matches were made at the epoch of the target catalogues by applying the *Gaia* proper motions to possible counterparts. The number of matches within 10″ are 103,317 objects from the Two Micron All-Sky Survey (hereafter 2MASS; Cutri et al. 2003; Skrutskie et al. 2006), 38,889 from the Sloan Digital Sky Survey Data Release 12 (SDSS; Abolfathi et al. 2018); 6,775 from the UKIRT Infrared Deep Sky Survey Galactic Clusters Survey (UKIDSS GCS; Lawrence et al. 2007); 107,009 from the Wide-field Infrared Survey Explorer (AllWISE; Cutri et al. 2014; Wright et al. 2010); and 118,144 from the first data release of the Pan-STARRS DR1 (Chambers et al. 2016; Kaiser et al. 2002). All surveys completely covered the total region apart from UKIDSS which only covered ∼20 square degrees centered on the Hyades nominal center.

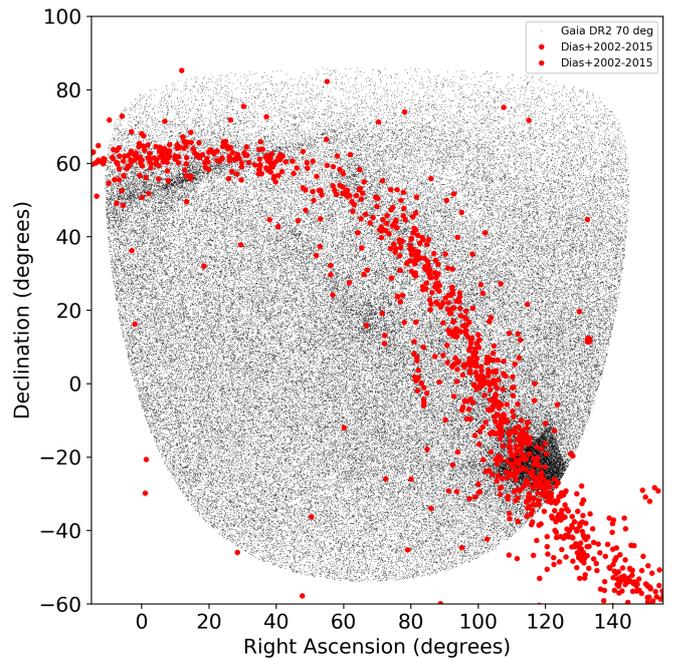

**Fig. 1.** Distribution of all sources in a radius of 70° from the cluster center (small black dots) along with open clusters close to galactic plane (Dias et al. 2002, 2006, 2018, 2014).

## 3. Previously known members

### 3.1. Compilation of known members

We compile a list of known members from different surveys published over the past decade which represent the most complete surveys in the Hyades. We started with the following sub-samples: 724 stars from Röser et al. (2011), 773 stars from Goldman et al. (2013) as well as 20 candidates from Dobbie et al. (2002). We also included 10 confirmed L dwarfs from Hogan et al. (2008) confirmed spectroscopically by Casewell et al. (2014) and Lodieu et al. (2014b), two mid-L dwarfs from Schneider et al. (2017), one L5 from Pérez-Garrido et al. (2017), and the two T dwarfs from Bouvier et al. (2008). After remov-ing common sources from these catalogues, we are left with 837 high-probability candidates with proper motion and photometry consistent with membership to the Hyades.

### 3.2. Cross-correlation with *Gaia* DR2

We cross-matched this list with *Gaia* DR2 with a matching radius of 3″, returning 825 sources equivalent to a recovery rate of 99%. The sources not recovered are mainly sub-stellar, including HyaL5 (Pérez-Garrido et al. 2017), the two mid-L dwarfs from Schneider et al. (2017), the two T dwarfs from Bouvier et al. (2008), Hya09 and Hya12 from Hogan et al. (2008), as well as four sources (ID = 30, 324, 651, 671)) from Goldman et al. (2013) and one (h7334b) from Dobbie et al. (2002). The 10 brightest Hyades members are also recorded in the *Gaia* DR2 catalogue. We conclude that *Gaia* is complete down to the hydrogen-burning limit but incomplete in the sub-stellar regime in the Hyades cluster because it is missing the seven aforementioned brown dwarfs. We do not know why the five targets above from Goldman et al. (2013) and Dobbie et al. (2002) are missing from DR2 (less than 1%). We checked that those objects are real on the Digital Sky Survey and 2MASS images although one of them (h7334b) appears faint, which may explain why it has no entry in *Gaia*. Three of them have proper motions quoted in Simbad and two of them (Goldman30, Goldman324) have entries in the *Gaia* DR2 catalogue without astrometry but effective temperature estimates and 2-parameter solutions. We also note that Goldman30 is classified as a M5 at 22 pc by Newton et al. (2015) and Goldman671 might belong to a young moving group (Gagné et al. 2015b).

We cross-correlated these 837 pre-*Gaia* members with our catalogue within 70° from the Hyades center and recovered 749 of these previously-known sources. The objects not recovered in our catalogue (837−749) have parallaxes less than 10 mas, which is our lower limit to create the 70° catalogue.

We have also cross-correlated our list of 749 pre-*Gaia* members with the catalogue of 515 Hyades sources published by Gaia Collaboration et al. (2018a), yielding 415 objects in common. We used the SourceID parameter for the cross-correlation to avoid any mis-matching based on coordinates. Therefore, we conclude that earlier surveys of the Hyades may have excluded about 20% of known members if the new candidates from Gaia Collaboration et al. (2018a) are confirmed as members. We note that Röser et al. (2011) estimated a level of contamination of the order of ~9% with 65 field stars in their sample of 724 Hyades candidates based on a control sample of stars with Hipparcos parallaxes and/or published radial velocities.

## 4. Ground-based parallaxes of Hyades L/T dwarfs

### 4.1. Targets

We selected the faintest Hyades members confirmed spectroscopically from three surveys. First, ten of the 12 ultracool dwarf candidates from Hogan et al. (2008), confirmed spectroscopically as late-M and early-L dwarfs by Casewell et al. (2014) and Lodieu et al. (2014b). Second, the recent L5 dwarf discovered by Pérez-Garrido et al. (2017) and confirmed as a sub-stellar member by Lodieu et al. (2018). Third, the two early-T dwarf members reported by Bouvier et al. (2008), the coolest members known to date.

We estimated the *Gaia* $G$-band magnitude of these sources with the equation $G − J \sim 0.244 \times \mathrm{SpT} − 12.633$, where SpT is the numerical counterpart of the spectral type with L0≡70 and T2≡82 as examples (Smart et al. 2017a).

We discarded the confirmed late-M and L dwarfs brighter than $G = 20.7$ mag because we predicted they would have *Gaia* astrometry. We programmed the remaining eight sources for astrometric follow-up with the infrared camera on the Liverpool telescope. This strategy has been proven to be effective because all the sources not included in our parallax program have *Gaia* distances (Table 1), except Hya04 (2MASS J04421859+1754373; Hogan et al. 2008) which has been included in the *Gaia* DR2 (ID = 3409343115420601728) with just positions; we expect it to have full astrometry in the next release.

### 4.2. Observations

We targeted the eight Hyades L and T dwarf member candidates with the infrared camera IO:I (Barnsley et al. 2016) on the 2-m robotic Liverpool telescope (Steele et al. 2004) over three semesters between August 2015 and January 2018 (CL15B06, CL16B03, CL17B01; PI Lodieu). We requested a seeing better than 1.5″ and an elevation on the sky higher than 30°. We also asked for a sky brightness better than the "dark+4 magnitudes" definition of the Liverpool telescope, which means that our program could be observed with bright moon and astronomical twilight in the worst case.

The IO:I instrument was installed on the Liverpool telescope in August 2015. It uses a 2048×2048 pixel Hawaii 2RG offering a field of view of 6.3 arcmin and a pixel scale of 0.18″. It is currently solely equipped with a *H*-band filter. The blue and red cut-offs at 1.4 and 1.8 microns are set by the detector and the filter, respectively.

To optimise the determination of the parallax from the ground, we designed the following strategy. We requested seven observations per semester for each target, distributed as follows: three points during morning and evening twilights spread over 1.5 months and separated by about two weeks to maximise the parallax factors and an additional point in the middle of the night to improve the proper motion solution. We repeated this strategy for three semesters over three years. However, the past two winter semesters have suffered from significantly bad weather, yielding a small number of points during the past two years. Nonetheless, we were able to collect between 10 and 17 points per object (Table A.1–A.8).

We employed the same dithering procedure for all targets: we used individual on-source integrations of 10s with a 9-point dither pattern to optimise the sky subtraction in the *H*-band. We repeated this sequence seven times for all objects, yielding a total on-source exposure time of 630s, except for the two faintest targets: the L5 dwarf (Pérez-Garrido et al. 2017) and CFHT-Hy-20 (Bouvier et al. 2008), which we repeated 10 times for a total time of 900s.

### 4.3. Data reduction

We downloaded directly from the archive the reduced images from the automatic pipeline designed for IO:I (Barnsley et al. 2016). We refer the reader to that paper for more detailed information. To summarise, the pipeline includes bias subtraction, non-linearity correction, flat fielding, bad pixel masking, and sky subtraction for each of the 9-point dither position resulting in seven or 10 repeated images with a world coordinate system incorporated.

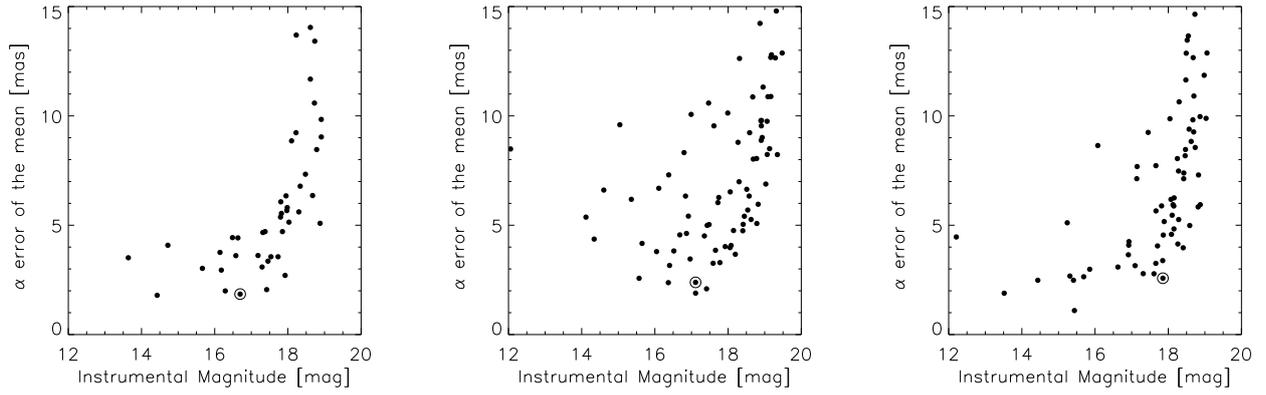

**Fig. 2.** Root-mean-square of the error of the mean between the first IO:I epoch and the remaining epochs as a function of instrumental *H*-band magnitude for three of the eight targets (Hya02, Hya09, and HyaL5 from left ot right) for which we determined ground-based distances. The circled objects are the targets.

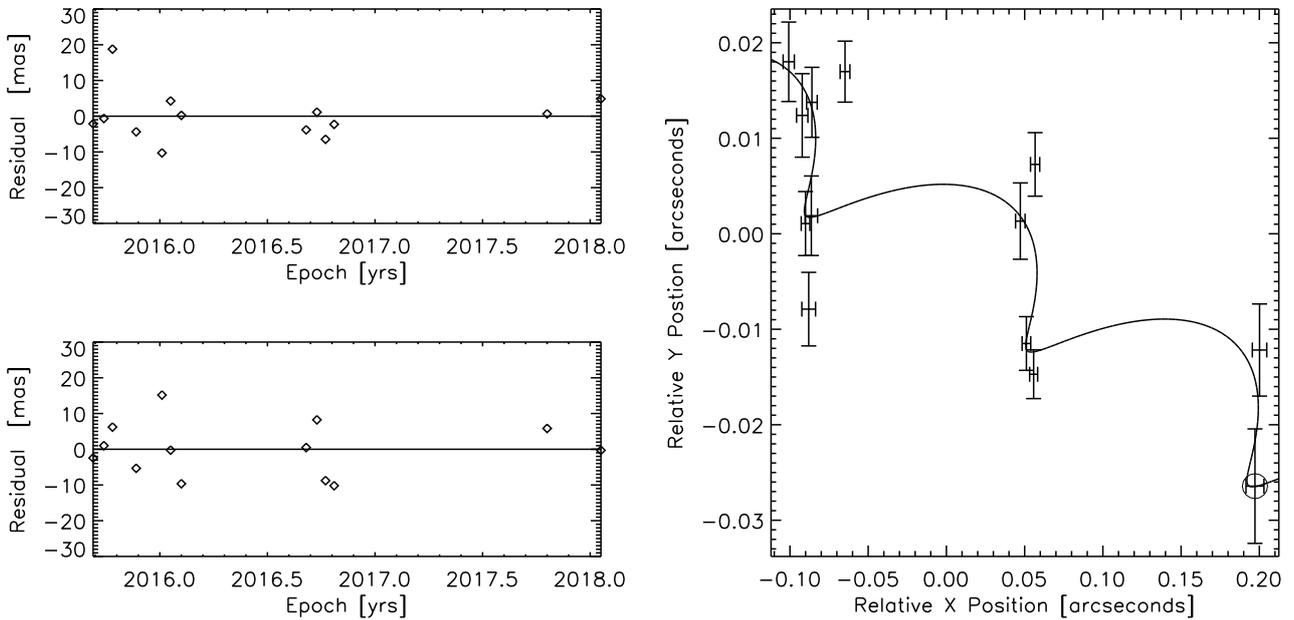

**Fig. 3.** *Left panels:* Residuals in mas as a function of epoch for right ascension (bottom) and declination (top). *Right panel:* Solution for the parallax determination. We show the example of CFHT-Hy-20, the other objects are plotted in Appendix A. The circled dot highlights the reference frame used for the determination of the parallax.

In a second step, we combined the seven or 10 repeated positions with the `imcombine` task under IRAF (Tody 1986, 1993) to create a final image for each target equivalent to the total on-source integrations quoted in Sect. 4.2. The stacking and offsetting of the individual repeats were performed within the task `imcombine` using the offsets from the header. We verified that the full-width-half-maximum on the final image was consistent with the range of seeing measured on individual repeats. We used the combined images to proceed with the centroiding procedure to derive trigonometric parallaxes (Fig. 2). The centroiding of all objects in the combined images was carried out using the Cambridge Astronomy Survey Unit's *imcore* maximum likelihood barycenter (CASUTOOLS; v 1.0.21; `casu.ast.cam.ac.uk`). Other packages were also tried but the centroiding from this package were the most robust and provided the most consistent floor to the precision as shown in Fig. 2.

### 4.4. Astrometric parameter determination

The astrometric reduction was carried out using the Torino Observatory Parallax program pipeline procedures and the reader is referred to Smart et al. (2003) for details and Smart et al. (2010) for some results. Here we just outline the main steps of the procedure. A base frame, observed on a night with good seeing, was selected and the measured x,y positions of all objects were transformed to a standard coordinate $\xi, \eta$ system determined from a gnomic projection of the *Gaia* DR2 objects in the frame. All subsequent frames were transformed to this standard coordinate system with a simple six constant linear astrometric fit using all common objects except the target. We then removed any frames that had an average reference star error larger than three times the mean error for all frames in either coordinate, or, had less than 12 stars in common with the base frame. This cleaning resulted in one frame being removed in the solutions of Hya10/Hya12/HyaL5 out of 12/17/12 frames, respectively, and two frames from the 15 in the Hya19 sequence.

**Table 1.** Hyades late-M and early-L confirmed spectroscopically with *Gaia* proper motions and distances. Note: Hya05 (M3) and Hya07 (M5) are not listed below because they were rejected as spectroscopic members of the Hyades based on their optical spectra (Lodieu et al. 2014b). The other L dwarfs are not in *Gaia* but listed in Table 2 with ground-based parallaxes.

| Name | RA | dec | SpT | SourceID | Plx | pmRA | pmDEC | *H* |
|------|-----|-----|-----|----------|-----|------|-------|-----|
| | hh:mm:ss.ss | dd:″:″.″ | | | mas | mas/yr | mas/yr | mag |
| Hya01 | 04:20:24.50 | +23:56:13.0 | M8.5 | 149089760932648448 | 22.6480±0.4575 | 130.178±0.876 | −28.598±0.660 | 13.85 |
| Hya03 | 04:10:24.01 | +14:59:10.3 | L0.5 | 3311691879984803072 | 17.3183±1.5853 | 107.981±2.747 | −11.185±2.612 | 14.78 |
| Hya04 | 04:42:18.59 | +17:54:37.3 | M9.5 | 3409343115420601728 | — | — | — | 14.97 |
| Hya06 | 04:22:05.22 | +13:58:47.3 | M9.5 | 3310992904122021120 | 18.2032±0.9181 | 89.391±1.612 | −17.657±1.307 | 14.81 |
| Hya08 | 04:58:45.75 | +12:12:34.1 | L0.5 | 3295377360811741184 | 24.2448±0.9774 | 85.748±2.536 | −16.008±1.265 | 14.55 |

**Table 2.** Hyades L/T members with ground-based parallaxes from the Liverpool telescope. We give the names, coordinates at epoch 2015.5 to be consistent with *Gaia* DR2 (Table 1), optical spectral types (Lodieu et al. 2014b; Martín et al. 2018), epochs of the base frame, the number of observations used in the solutions ($N_{obs}$) with the numbers of reference stars entering the fit ($N_{star}$), the corrections from relative to absolute parallax (COR), the proper motions, the baselines of the observational sequence (ΔT).

| Name | $\alpha$ (2015.5) | $\delta$ (2015.5) | SpT | Baseframe | $N_{obs}$,$N_{star}$ | $\varpi_{abs}$ | COR | $\mu_\alpha \cos \delta$ | $\mu_\delta$ | ΔT |
|------|------|------|-----|-----------|-----------|------|-----|------|------|-----|
| | hh:mm:ss.sss | dd:mm:ss.sss | | year | | mas | mas | mas/yr | mas/yr | yr |
| Hya02 | 03:52:46.433 | +21:12:32.805 | L1.5 | 2016.7913 | 11, 36 | 17.7±2.0 | 0.96 | 116.4±2.0 | −26.9±1.5 | 2.38 |
| Hya10 | 04:17:33.988 | +14:30:15.360 | L2.0 | 2015.7622 | 11, 36 | 28.5±3.9 | 0.96 | 120.2±3.6 | −12.2±5.5 | 2.39 |
| Hya11 | 03:55:42.143 | +22:57:01.073 | L3.5 | 2016.7371 | 10, 47 | 39.1±16.3 | 0.96 | 138.1±13.3 | −19.4±9.0 | 2.37 |
| Hya12 | 04:35:43.043 | +13:23:44.877 | L3.0 | 2016.8054 | 16, 63 | 24.1±2.1 | 0.92 | 100.2±1.9 | −15.1±2.0 | 2.33 |
| Hya19 | 04:46:35.444 | +14:51:25.951 | L4.0 | 2016.7342 | 13, 80 | 20.6±2.5 | 0.82 | 76.3±2.9 | −17.7±1.5 | 2.12 |
| HyaL5 | 04:18:35.011 | +21:31:26.788 | L5.0 | 2016.7315 | 11, 67 | 25.8±2.9 | 0.81 | 141.5±2.7 | −45.7±2.3 | 2.35 |
| CFHT-Hy-20 | 04:30:38.887 | +13:09:56.636 | T2.0 | 2018.0531 | 13, 48 | 30.8±3.0 | 0.77 | 141.3±2.9 | −14.5±3.2 | 2.37 |
| CFHT-Hy-21 | 04:29:22.869 | +15:35:29.842 | T1.0 | 2015.7815 | 10, 72 | 33.5±12.7 | 0.79 | 82.1±9.8 | −15.5±8.6 | 2.38 |

Since the target is not used in the fit, its positional change is a reflection of its parallax and proper motion. We fit a simple five parameter model to this positional change, and that of all the other objects in the field, to find their astrometric parameters implicitly assuming that all objects are single. We then iterate this procedure where, in addition to removing frames as described above, we also remove objects with large errors over the sequence from the subset used to astrometrically align frames. Finally, for the target we removed any observations where the combined residuals of the two coordinates is greater than three times the sigma of the whole solution. No attempt was made to improve the astrometric fit by assuming a binary system because the length of the observational sequence and small number of observations would not support such a fit.

The solutions were tested for robustness using bootstrap-like testing where we iterate through the sequence selecting different frames as the base frame thus computing many solutions that incorporate varied sets of reference stars and starting from different dates. We create the subset of all solutions with: (i) a parallax within 1$\sigma$ of the median solution; (ii) the number of included observations in the top 10%; and (iii) at least 12 reference stars in common to all frames. From this subset, for this publication, we have selected the one with the smallest error. More than 90% of the solutions were within 1$\sigma$ of the published solution.

To the relative parallaxes we add a correction to find astrophysically useful absolute parallaxes. This correction is estimated from the difference of the median *Gaia* DR2 parallaxes of the common reference stars to the measured median parallax calculated from the observations. As can be seen from Table 2, this correction is always less than 1 mas and we added 20% of the correction to the formal parallax uncertainty in quadrature to obtain the quoted errors. The results are summarised in Tables 1–2 and in Fig. 3 we show the on-sky motion and the residuals for target CFHT-Hy-20 along with the solution. The epochs of observations and plots of the solution and residuals for the other targets are included in Tables A.1–A.8 and Figs. A.1–A.4 in Appendix A.

This is the first time that parallaxes have been determined with IO:I on the Liverpool robotic telescope and demonstrates that it is possible to derive useful ground-based parallaxes up to ∼50 pc. There remains an important sample of very cool nearby objects that are too faint for *Gaia* but possible on this telescope/instrument combination.

## 5. Selection of Hyades member candidates

In this section, we implemented the kinematic procedure described by Perryman et al. (1998), whose technique was originally described in Jones (1971). This method determines the barycenter of the cluster and identifies potential members based on their velocities in space (Sect. 5.1).

We decided to use this method because it was specifically developed for the Hipparcos satellite, whose astrometry offers much higher accuracy than ground-based surveys. The advent of *Gaia* provides even more accurate parallaxes and proper motions for a significantly larger number of stars in the Galaxy.

### 5.1. The kinematic method

We applied the kinematic procedure of Perryman et al. (1998) to the sample of 126,144 objects located within a radius of 70° from the cluster center at (RA,Dec) = (67,16)°. This method has been successfully applied to the Hyades using Hipparcos (de

**Table 3.** Positional and kinematics data of the Hyades cluster. First row shows the data obtained with the preliminary list of 154 Hyades members. Second (third) row display the results after recalculating cluster center and velocity with objects from the final list closer than 10 pc (20 pc) to the cluster center.

| selection | $N$ | $\mathbf{b}_c$(pc) | | | $\mathbf{v}_c$(km s$^{-1}$) | | | Distance | Velocity |
|---|---|---|---|---|---|---|---|---|---|
| | | $b_x$ | $b_y$ | $b_z$ | $v_x$ | $v_y$ | $v_z$ | pc | km s$^{-1}$ |
| Pre | 154 | $-44.49\pm0.16$ | $0.18\pm0.10$ | $-17.10\pm0.08$ | $-42.23\pm0.09$ | $-19.23\pm0.03$ | $-1.18\pm0.05$ | $47.66\pm0.18$ | $46.42\pm0.10$ |
| $r < 10pc$ | 122 | $-43.83\pm0.18$ | $0.42\pm0.11$ | $-17.05\pm0.09$ | $-42.14\pm0.11$ | $-19.26\pm0.04$ | $-1.12\pm0.05$ | $47.03\pm0.20$ | $46.34\pm0.12$ |
| $r < 20pc$ | 168 | $-44.45\pm0.15$ | $0.26\pm0.10$ | $-17.16\pm0.08$ | $-42.19\pm0.10$ | $-19.25\pm0.03$ | $-1.14\pm0.05$ | $47.65\pm0.17$ | $46.39\pm0.11$ |

**Table 4.** Summary of numbers of members in our work and from the various cross-matches with catalogues from previous studies (Sect. 5.2).

| Method | Members | Comments |
|---|---|---|
| This paper | 85, 381, 568, 710 | within 3.1, 9, 18, 30 pc |
| Recovered in Perryman et al. (1998) | 129 out of 192 | Memb = 1 from Perryman; 127 within 30 pc |
| Recovered in Perryman et al. (1998) | None | Memb = ? or 0 from Perryman |
| Recovered in Gaia Collaboration et al. (2018a) | 503 out of 515 | all within 20 pc |
| Recovered in Reino et al. (2018) | 169 out of 187 | *Gaia* DR1; 159 within 30 pc |
| Recovered among pre-*Gaia* known members | 70, 306, 443, 518 | within 3.1, 9, 18, 30 pc from 749 |
| Recovered in BANYAN | 85, 376, 469, 484 | within 3.1, 9, 18, 30 pc (568 candidates using BANYAN) |

Bruijne et al. 2001; Perryman et al. 1998) and *Gaia* DR1 (Reino et al. 2018).

We implemented the Bayesian method using equations 17 and 18 of Luri et al. (2018) to transform *Gaia* parallaxes into distances. To determine the membership probability of all objects in the 70° area, we calculate the space velocity of the Hyades cluster following Perryman et al. (1998). For the sake of clarity we sketch this method below. The *Gaia* mission provides high quality parallax measurements ($\pi$), proper motions ($\mu_\alpha\cos\delta, \mu_\delta$), and radial velocities ($V_R$); the latter only for the brightest sources ($G \sim 5$–13 mag). In a first step we calculate the cluster barycenter ($\mathbf{b}_c$) and space velocity ($\mathbf{v}_c = v_x, v_y, v_z$) using a preliminary set of Hyades members selected by Gaia Collaboration et al. (2018a). These quantities are calculated using the standard expressions:

$$\mathbf{b}_c = \frac{\sum m_i \mathbf{b}_i}{\sum m_i} \quad \mathbf{v}_c = \frac{\sum m_i \mathbf{v}_i}{\sum m_i} \tag{1}$$

where $\mathbf{b}_i = (d_i \cos\alpha_i \cos\delta_i, d_i \sin\alpha_i \cos\delta_i, d_i \sin\delta_i)$ is the positional vector for object $i$, with equatorial coordinates $\alpha_i$ and $\delta_i$ and located at a distance $d_i$ (measured in pc). The mass $m_i$ is estimated using predicted magnitudes and $G - J$ colours from the Padova and BT-Settl models (see Sect. 7.2 for more details). The mass is used in the calculation of the barycenter but as a weight. Assuming that binaries are distributed isotropically with respect to the center then the barycenter should not change. We checked that this is the case assuming that all stars have a mass of 1 M$_\odot$, which would largely account for binaries. This fact is in agreement with the conclusions of Perryman et al. (1998) and Reino et al. (2018) who concluded that "these results are rather insensitive to the weighting scheme adopted". In conclusion, binaries do impact on the total mass and the mass function (Sect. 7.2) but not the barycenter. The velocity vectors ($\mathbf{v}_i$) are calculated using the object transverse and radial velocities with the equation:

$$\begin{pmatrix} v_{ix} \\ v_{iy} \\ v_{iz} \end{pmatrix} = \mathbf{R}_i \begin{pmatrix} V_{i\alpha*} \\ V_{i\delta} \\ V_{iR} \end{pmatrix}, \tag{2}$$

where $V_{i\alpha*} = \mu_{i\alpha*}A_v/\pi_i$, $V_{i\delta} = \mu_{i\delta}A_v/\pi_i$, and $V_{iR}$ are the observed transverse and radial velocities, with $A_v = 4.74047$ km yr s$^{-1}$.

The matrix $\mathbf{R}_i$ is given by:

$$\mathbf{R}_i = \begin{pmatrix} -\sin\alpha_i & -\sin\delta_i\cos\alpha_i & \cos\delta_i\cos\alpha_i \\ \cos\alpha & -\sin\delta_i\sin\alpha_i & \cos\delta_i\sin\alpha_i \\ 0 & \cos\delta_i & \sin\delta_i \end{pmatrix} \tag{3}$$

In a second step, we use $\mathbf{v}_c$ to select objects in the *Gaia* database with motions consistent with the cluster. To carry out this task, we estimate the expected transverse and radial velocities at the position of each object:

$$\begin{pmatrix} V_{i\alpha*}^e \\ V_{i\delta}^e \\ V_{iR}^e \end{pmatrix} = \mathbf{R}_i^{-1} \begin{pmatrix} v_x \\ v_y \\ v_z \end{pmatrix} \tag{4}$$

where $\mathbf{R}_i^{-1}$ is actually the transpose matrix of $\mathbf{R}_i$ from Eq. 3. We define the vector $\mathbf{z}_i$ as the difference vector between the observed and expected transverse and radial velocities. We need to calculate two covariance matrices, one associated to the observed transverse/radial velocities and the other associated to the expected ones (see detailed explanation in Perryman et al. 1998). Assuming that these velocities are statistically independent, the sum of their two covariance matrices $\Sigma$ describes the combined confidence region and the parameter:

$$c = \mathbf{z}^T \Sigma^{-1} \mathbf{z} \tag{5}$$

is a $\chi^2$ statistical test with three degrees of freedom (DOF). As not all the objects in *Gaia* have radial velocities we adapted the method to work also with transverse velocities only. For sources without radial velocity the test has only two DOF. We select as good candidates those objects with velocities within $4.4171\sigma$ of common cluster motion, which correspond to a p value of 0.99999. Thus for three DOF (objects with radial velocity measured), all sources with $c < 25.9$ are considered as Hyades members, while for objects without radial velocity in *Gaia* data, i.e. two DOF statistic, the threshold is 23.03.

From the preliminary list of Hyades members from Gaia Collaboration et al. (2018a) we calculate $\mathbf{b}_c$ and $\mathbf{v}_c$, as explained above, and check whether all sources in that list can be considered as good members of the cluster using the value of the $c$ parameter (Eq. 5). Discarding those with values of $c$ larger than the chosen threshold and then recalculating new values for $\mathbf{b}_c$ and $\mathbf{v}_c$. This procedure is repeated until no further objects are discarded. At the end of this process we have 154 objects and the final values of $\mathbf{v}_c$ are employed to assess which objects from *Gaia* catalogue can be considered as bona-fide Hyades members. In Table 3 we list the barycenter and cluster velocity in Galactic coordinates. We apply this procedure to the list of 126,144 sources in the Hyades region, finding 1764 objects with common motions to the cluster. This final list is then used to re-estimate $\mathbf{b}_c$ and $\mathbf{v}_c$ (Table 3). The cluster center velocity obtained when we use objects located in a radius of 10 pc from the cluster center is employed to carry out a second iteration with the final list of 1764 objects checking that all of them continue to be considered as good Hyades members.

As discussed at length in Röser et al. (2011), most of the members of the cluster lie within its tidal radius. In the case of the Hyades, all objects within 9 pc are most likely bound while the candidates up to 18 pc most likely belong to its halo. Kinematic candidates in the 18–30 pc might belong to the Hyades moving group (Boss 1908; Eggen 1958; Famaey et al. 2007; Zuckerman & Song 2004) but we expect a significant level of contamination. We do not consider the candidates beyond the 30 pc limit in this work (710 sources). We limit our analysis to the core (3.1 pc), tidal radius (9 pc), halo (18 pc), and up to 30 pc from the cluster center in 3D space to allow for direct comparison with the work of Röser et al. (2011). The kinematic procedure returned 85, 381, 568, and 710 sources within 3.1, 9, 18, and 30 pc, respectively (Fig. 4–5). The large area at around RA = 90–100° and declinations around zero containing candidates further than 30 pc from the center of the Hyades corresponds to clusters in the galactic plane, and possibly to Platais 6, whose extension on the sky is estimated to be 250 arcmin (Dias et al. 2002, 2006, 2018, 2014).

We display several colour-magnitude diagrams in Appendix B, showing combinations of *Gaia* magnitudes with other large-scale surveys (Figs. B.1–B.3). We display all candidates identified in this work as black dots and highlight the members located within 3.1, 9, 18, and 30 pc with yellow, blue, green, and red symbols, respectively. The sequences drawn in those colour-magnitude diagram represent benchmark sequences at an age of 650 Myr, which are key to compare with other clusters and members of moving groups.

### 5.2. Comparison with previous Hyades surveys

#### 5.2.1. Hipparcos catalogue from Perryman et al. (1998)

We cross-correlated our sample of Hyades candidate members with the Hipparcos catalogue (Perryman et al. 1998), which is divided up into three groups: 192 members (Memb = 1), 20 probable members (Memb = ?), and 60 non-members (Memb = 0). The *Gaia* DR2 catalogue contains 183 of the 192 members (>95% completeness) but only 13 of the 69 probable members (18.8%) and 31 of the 60 non-members (51.7%). We recovered 129 sources of the 192 objects classified as members (Memb = 1 in their table), see Table 4 in our full list, including 127 being within 30 pc. The remaining 63 sources do not satisfy the criteria of the kinematic method because they have "c" indices larger than 25.9 or 23.03 depending on the availability of *Gaia* radial velocity.

None of the other candidates reported by Perryman et al. (1998) as probable members or likely non-members in their table have counterparts in our catalogue of candidates within 70° from the cluster center.

We expect some difference because the cluster parameters derived from *Gaia* DR2 are slightly different from those inferred by Hipparcos: from our 10-pc sample, we find that a mean distance and velocity of the cluster is 47.03±0.20 pc and 46.38±0.12 km s$^{-1}$, respectively, compared to 46.34±0.27 pc and 45.93±0.23 km s$^{-1}$ for the Hipparcos 10-pc sample (Table 3 of Perryman et al. 1998). We note that the global *Gaia* zero point is −0.03 mas but it varies upon position in the sky by 0.15 to −0.15 mas (Lindegren et al. 2018). This correction is negligible compared to the average parallax of the Hyades (∼21–22 mas). The *Gaia*–Hipparcos offset is −0.118 mas (Arenou et al. 2018), which means that the distance of the Hyades from Hipparcos put on the *Gaia* system would equate into 46.595 pc, close to our estimate using the 10 pc sample. This is the distance we adopt in this work. The distance from Gaia Collaboration et al. (2018a) makes use of all kinematic members up to 16 pc (Section 4 in Gaia Collaboration et al. 2017) and should therefore be closer to our 20-pc estimate, which is indeed the case within 1σ. This difference will lead to some variation in the numbers of kinematic members.

#### 5.2.2. *Gaia* DR2 catalogue (Gaia Collaboration et al. 2018a)

We cross-correlated our sample with the list of 515 member candidates from Gaia Collaboration et al. (2018a). The matching was based on the `SourceID` parameter to avoid any mis-matches due to positional matching problems. We recovered 503 out of 515 objects that lie in a radius of approximately 16 pc, the limit set intrinsically by the procedure described in Gaia Collaboration et al. (2018a). We find that only a small number of sources of Gaia Collaboration et al. (2018a) are not in our sample of member candidates. The small difference in numbers is most likely due to the divergence between our space velocity and distance estimates. Furthermore, Gaia Collaboration et al. (2018a) clearly stated in their Sect. 2 that they "selected the most precise data, without trying to reach completeness". This is particularly the case in the low-mass and sub-stellar regimes where *Gaia* gives larger astrometric errors but remains nonetheless reliable. We noticed this effect when comparing their candidates with our sample in the various colour-magnitude diagrams presented in this work.

#### 5.2.3. TGAS catalogue from Reino et al. (2018)

We collected the *Gaia* DR2 `SourceID` of the 251 candidates identified by Reino et al. (2018) based on Tycho-*Gaia* DR1 catalogue. Reino et al. (2018) kept 187 out of 251 candidates as bona-fide members while the other were rejected on the basis of their modelled parallaxes and standard errors. We found that 242 of the 251 have *Gaia* DR2 counterparts. We checked that the nine missing objects have entries in the second release of *Gaia* without parallax and proper motion. We recovered 169 as potential members of the Hyades, including 159 within 30 pc from the center (Table 4).

#### 5.2.4. Known members

We cross-correlated our sample with the list of 749 known members with *Gaia* DR2 parallaxes and located within 70° of the cluster. We conducted the cross-correlation with the `SourceID` keyword as before. In the full 70°, we retrieved 527 pre-*Gaia* known members. We recover eight of the 10 brightest Hyades members, except $\Theta^1$ Tauri (G5III; c = 398.325; Keenan & Mc-

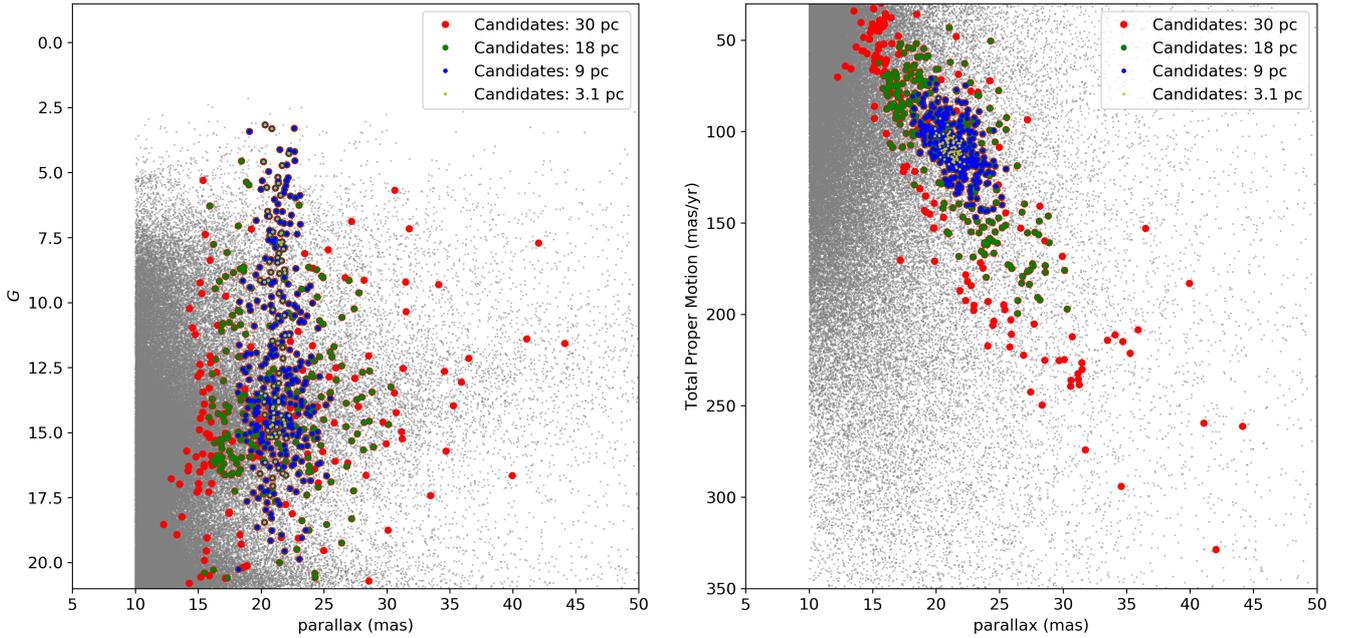

**Fig. 4.** *Left:* Parallax as a function of the *G*-band magnitudes from *Gaia* DR2 for candidates within radii of 3.1 pc (yellow), 9 pc (blue), 18 pc (green), and 30 pc (red) from the cluster center. *Right:* Parallax as a function of the total proper motion from *Gaia* DR2. We added as small grey dots the full *Gaia* catalogue over the 70° radius.

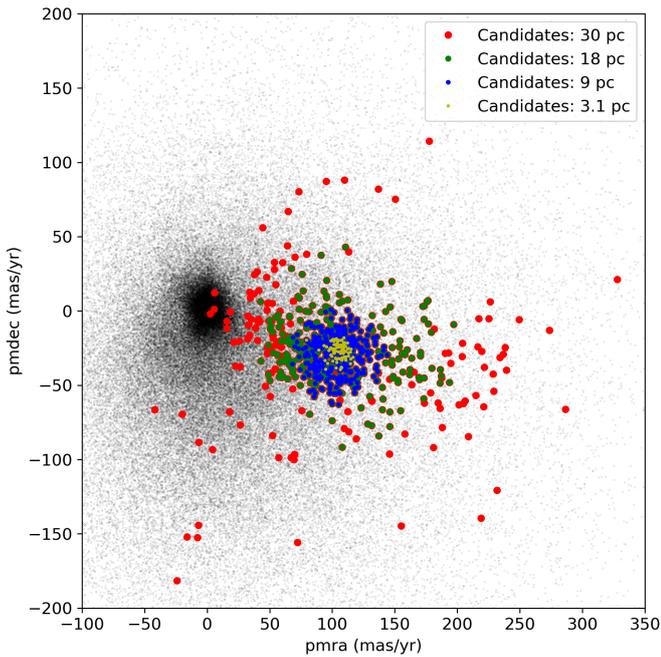

**Fig. 5.** Vector point diagram with proper motions in right ascension and declination for all sources in the full 70° catalogue (grey). We over-plot kinematic candidate members belonging to the Hyades and located within radii of 3.1 pc (yellow), 9 pc (blue), 18 pc (green), and 30 pc (red).

Neil 1989) and 71 Tauri (F0V; $c = 85.39$), whose *Gaia* parallaxes and proper motions confirm their membership. The former is a known spectroscopic binary and the latter is a known variable star. These eight bright members are the brightest sources in the colour-magnitude diagrams showing *Gaia* photometry (e.g. Fig. 6). All these sources lie within 6 pc from the cluster cen-ter. Therefore, $749 - 527 = 222$ previously-known members are rejected by the kinematic method because their "c" indices are larger than the maximum values allowed for membership even though their proper motions, parallaxes and magnitudes maybe consistent with other Hyades members. We conclude that the level of contamination from earlier surveys using the convergent point method is about 30%, consistent with ground-based photometric surveys in the Pleiades (Bouvier et al. 1998; Moraux et al. 2001) and Alpha Persei (Barrado y Navascués et al. 2002; Lodieu et al. 2005). We recovered 140, 312, 446, and 522 pre-*Gaia* known members within 3.1, 9, 18, 30 pc from the cluster center, respectively (Table 4). Hence, the level of completeness of previous surveys is decreasing with larger radii, from 80.9% in the 3.1 pc radius to 73.6% in the halo.

### 5.3. Comparison with BANYAN

Malo et al. (2013) developed a new method based on a Bayesian analysis to identify new members of nearby young kinematic groups and assess their membership probability. This method, nicknamed BANYAN (Bayesian Analysis for Nearby Young AssociatioNs) was later updated by Gagné et al. (2014) and further improved by Gagné et al. (2018a) to model the 6D space with multivariate Gaussians to take advantage of *Gaia*. This algorithm has been successfully employed to revise the membership of known members of young moving groups and also identify new candidates in a series of papers

The new BANYAN Σ algorithm is freely available[1] and includes the Hyades in the list of nearby associations. We used the IDL version which allowed us to calculate probabilities using proper motions, parallaxes and radial velocities or subsets of those parameters. We did not incorporate any extra constraints, for example indicators of youth. We ran this algorithm on the full 70° catalogue (126,144 objects), and 568 objects were returned with a most probable Bayesian hypothesis of being a Hyades member. The cluster parameters given in Table 9 of Gagné et al. (2018a) are the main parameters derived by Perryman et al. (1998), which we have now improved with the higher accuracy provided by *Gaia* DR2.

---

[1] www.exoplanetes.umontreal.ca/banyan/banyansigma.php

**Table 5.** Hyades WDs in *Gaia* DR2: single WDs are at the top and binary WDs at the bottom.

| SourceID | Name | RA | dec | Plx | pmRA | pmDEC | Bp | Rp | G |
|---|---|---|---|---|---|---|---|---|---|
| | | deg | deg | mas | mas/yr | mas/yr | mag | mag | mag |
| 45980377978968064 | EGGR29 | 62.3709338 | 17.1316806 | 19.9402±0.0931 | 111.4542±0.2682 | −22.0215±0.1464 | 15.302 | 15.460 | 15.378 |
| 3294248609046258048 | HZ14 | 70.2576501 | 10.9943984 | 20.2471±0.0510 | 91.3544±0.0937 | −10.6426±0.0497 | 13.669 | 14.108 | 13.864 |
| 3306722607119077120 | HZ7 | 68.4377818 | 12.7111484 | 21.1400±0.0616 | 99.0219±0.1237 | −14.3138±0.0711 | 14.122 | 14.437 | 14.262 |
| 3308403897837092992 | EGGR316 | 70.0999149 | 13.9793215 | 21.7321±0.0563 | 95.2761±0.1097 | −20.6871±0.0512 | 14.810 | 15.006 | 14.932 |
| 218783542413339648 | GD52 | 58.0005259 | 34.1241725 | 23.5584±0.0457 | 145.9926±0.0980 | −77.6952±0.0668 | 15.153 | 15.257 | 15.202 |
| 3313714023603261568 | EEGR37 | 67.1646712 | 16.9699097 | 20.8952±0.0567 | 102.6919±0.1147 | −26.8853±0.0681 | 13.913 | 14.288 | 14.075 |
| 3313606340183243136 | EGGR36 | 65.9825865 | 16.3540809 | 22.2272±0.0519 | 114.4115±0.1034 | −27.7153±0.0792 | 14.219 | 14.505 | 14.347 |
| 39305036729495936 | HG7−85 | 60.9260025 | 14.9912221 | 24.0527±0.0541 | 141.1900±0.1073 | −24.0698±0.0843 | 14.984 | 15.112 | 15.039 |
| 3302846072717868416 | HZ4 | 58.8423743 | 9.7883468 | 28.5890±0.0536 | 173.2722±0.1072 | −5.5099±0.0788 | 14.511 | 14.644 | 14.564 |
| 43789772861265792 | V471-Tau | 57.6046015 | 17.2464125 | 20.9569±0.0440 | 127.4578±0.0956 | −22.4774±0.0643 | 9.678 | 8.591 | 9.200 |
| 3310615565476268032 | HD27483 | 65.2201424 | 13.8643726 | 21.0518±0.0769 | 106.9751±0.1947 | −12.0173±0.1481 | 6.284 | 5.683 | 6.033 |
| 3314232855652895104 | HZ9 | 68.0994836 | 17.7505941 | 23.2618±0.0373 | 109.8651±0.0661 | −34.5059±0.0419 | 14.030 | 12.445 | 13.362 |
| 3311810043124387712 | LP474−185 | 63.4688348 | 15.3649605 | 23.6172±0.5733 | 116.2583±1.0037 | −24.2140±0.7910 | 15.417 | 12.495 | 14.257 |

**Table 6.** Derived parameters for the eight single DA WD members in the Hyades, considered for the age determination.

| SourceID | $M_{WD}$ | $M_{MS}$ | Time$_{WD}$ | Time$_{MS}$ | Time$_{total}$ | SpT | $T_{eff}$ | log $g$ |
|---|---|---|---|---|---|---|---|---|
| | $M_\odot$ | $M_\odot$ | Myr | Myr | Myr | | K | dex |
| 45980377978968064 | 0.826 | $3.823^{+0.637}_{-0.478}$ | 355.0 | $236.8^{+107.2}_{-80.2}$ | $591.8^{+107.2}_{-80.2}$ | DA3.2 | 15810±288 | 8.38±0.05 |
| 3294248609046258048 | 0.708 | $2.9657^{+0.412}_{-0.343}$ | 20.4 | $487.8^{+272.6}_{-152.7}$ | $508.2^{+272.6}_{-152.7}$ | DA1.8 | 27540±403 | 8.15±0.05 |
| 3306722607119077120 | 0.666 | $2.673^{+0.349}_{-0.299}$ | 78.8 | $664.7^{+272.6}_{-203.9}$ | $743.4^{+272.6}_{-203.9}$ | DA2.3 | 21890±346 | 8.11±0.05 |
| 218783542413339648 | 0.838 | $4.003^{+0.692}_{-0.508}$ | 475.2 | $208.5^{+95.2}_{-71.3}$ | $683.7^{+95.2}_{-71.3}$ | DA3.4 | 14820±350 | 8.31±0.05 |
| 3313714023603261568 | 0.691 | $2.863^{+0.388}_{-0.337}$ | 44.8 | $542.8^{+232.9}_{-169.8}$ | $587.6^{+232.9}_{-169.8}$ | DA2.0 | 25130±381 | 8.12±0.05 |
| 3313606340183243136 | 0.693 | $2.879^{+0.361}_{-0.330}$ | 112.8 | $533.7^{+229.8}_{-167.0}$ | $646.6^{+229.8}_{-167.0}$ | DA2.5 | 20010±315 | 8.13±0.05 |
| 39305036729495936 | 0.816 | $3.693^{+0.598}_{-0.456}$ | 400.4 | $260.2^{+111.7}_{-87.0}$ | $660.6^{+111.7}_{-87.0}$ | DA | 15131±209 | 8.48±0.02 |
| 3302846072717868416 | 0.780 | $3.373^{+0.510}_{-0.405}$ | 359.2 | $336.1^{+150.0}_{-109.3}$ | $695.2^{+150.0}_{-109.3}$ | DA3.4 | 14670±377 | 8.30±0.05 |

We cross-correlated our full sample of 1764 sources with the 568 candidates from BANYAN and found 484 objects in common. Limiting the analysis to the radii of 3.1, 9, 18, and 30 pc from the cluster center, we have 85, 376, 469, and 484 in common with BANYAN, respectively. The difference between 489 and 484 comes from source located at 30 pc. The remaining 568−484 = 84 objects from BANYAN are not in our list. About a quarter of these sources have right ascension below 60°, lying in the tail of members identified by BANYAN (Table 4). The differing candidate lists are most likely the result of the very different membership allocation procedures and also the different distance and space velocities between Hipparcos/BANYAN (Gagné et al. 2018a; Perryman et al. 1998) and this work.

## 6. White dwarfs

One striking feature in the Hertzsprung-Russell diagrams displayed in Figure 17 of Gaia Collaboration et al. (2018a) is the presence of white dwarf (WD) members at around $Bp − Rp < 0$ and $G \sim 14–16$ mag. The existence of WDs in the Hyades has been known since the discovery of 10 WDs by von Hippel (1998). Seven of these "classical" WD are single while three are binaries (Table 5) and one of them a doubtful member due to its distance and tangential velocity (Weidemann et al. 1992). Schilbach & Röser (2012) presented a compendium of WD candidates in the Hyades, including the 10 classical WDs (von Hippel 1998) and three new sub-groups: one with 12 new potential Hyades WDs (running number from 11 to 22) and two additional ones with most likely non-members (running number from 23 to 37).

We cross-matched this list of 37 WD candidates from Schilbach & Röser (2012) with our *Gaia* DR2 and recovered 13 WDs previously published in the literature (Table 5). All the 10 classical WDs (seven singles and three binaries) from von Hippel (1998) are confirmed as members based on *Gaia* parallaxes and proper motions and common to the sequence of WDs from Gaia Collaboration et al. (2018a). We note that the three classical WD binaries are unresolved in *Gaia* because they have orbital periods less than about 3 days. Three other WDs from the *Gaia* sample are common to the sub-group of new WD members in Table 1 of Schilbach & Röser (2012). The remaining 37−15 = 22 candidates in Schilbach & Röser (2012) are rejected as members based on their parallaxes and proper motions. All three brightest WDs are binaries and appear much redder in the *Gaia* colours (bottom panels in Fig. 6). The WD+dM binary (EGGR 38; HZ9) lies between the WD sequence and the cluster main sequence in the $(Bp − Rp, Rp)$ colour-magnitude diagram. The other two classical binaries lie on the cluster main-sequence: one is a known WD+F6 binary (HR 1358; HD 27483) and the other one is the known eclipsing binary of Algol type (WD+K2) V 471 Tau (Fig. 6). We find one more binary, LP 474-185 Schilbach & Röser (2012), based on *Gaia* photometry.

Table 5 lists all the binary and single WDs belonging to the Hyades. Of the nine single stars in Table 5, we considered only eight with a pure hydrogen atmosphere (DA spectral class) to determine the cluster age. The other one, EGGR 316, was excluded because it is a DBA with mixed H/He atmosphere (Bergeron et al. 2011). In Fig. 7, we compare the position of the eight single DA WDs in the $(G_{BP}-G_{RP}, M_G)$ colour-absolute magnitude diagram with the cooling tracks of DA white dwarfs with H-thick envelopes ($M_H/M_\star=10^{-4}$) from Bergeron et al. (2011)[2]. From cubic spline interpolation of the models, we determined the basic physical parameters of each WD: effective temperature, surface gravity, mass and cooling age. After having verified that the effective temperatures and surface gravities (log $g$) were in good agreement with the known spectroscopic values from the literature, we used the initial-to-final mass relation (IFMR) of El-Badry et al. (2018) to derive the mass of each WD progenitor. Then, for each WD progenitor, we used the Padova evolutionary models of massive stars (Bressan et al. 2012)[3] with nearly solar abundances (Z=0.017, Y=0.279) to compute the time needed to evolve from the pre-MS to the first thermal pulse in the asymptotic giant branch.

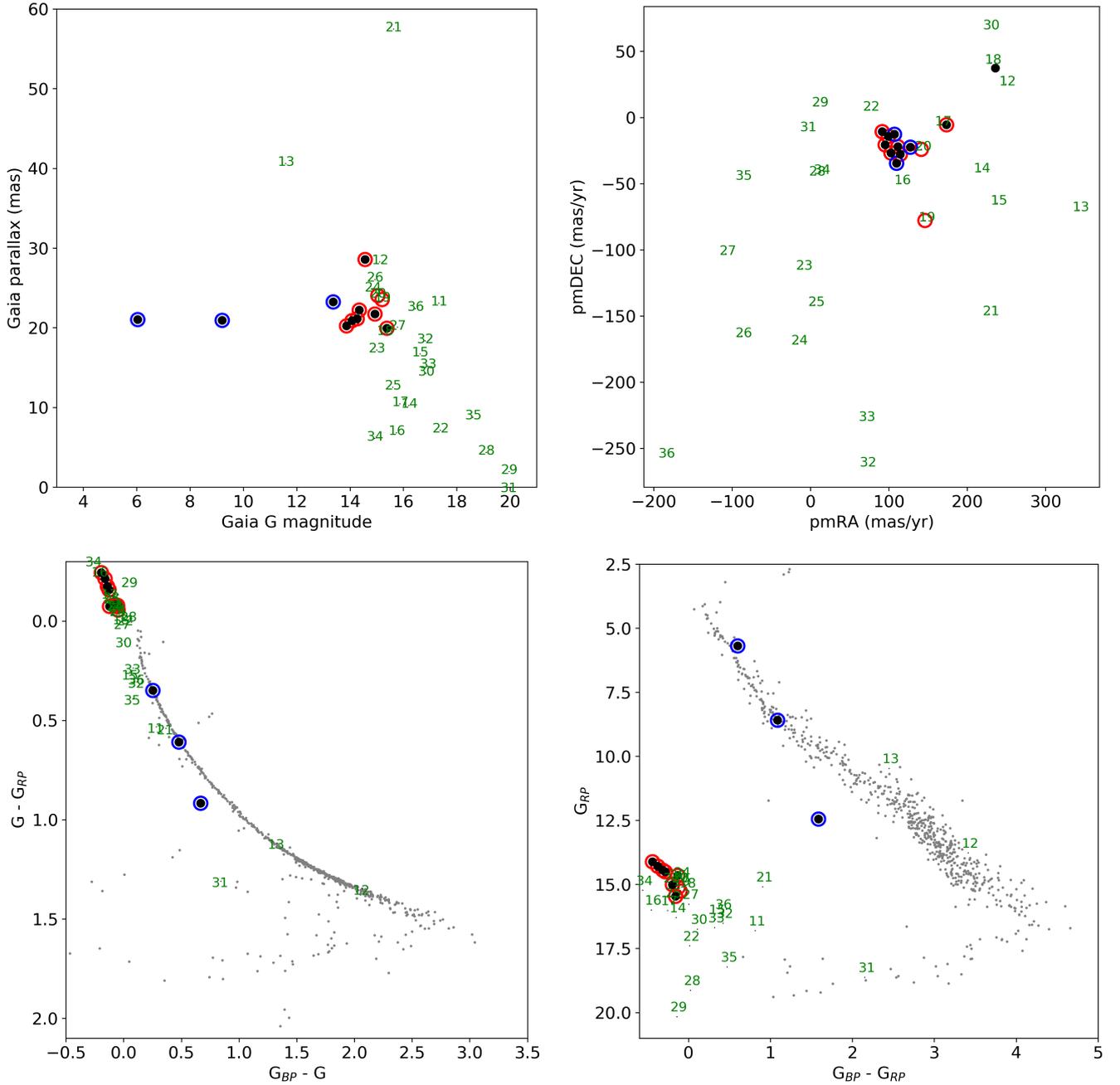

**Fig. 6.** *Top left:* Parallaxes of Hyades WD candidates as a function of *Gaia* magnitude. *Top right:* Vector point diagrams for Hyades WD candidates. *Bottom left:* $(G_{BP} - G, G - G_{RP})$ colour-colour magnitude of all candidates in our 30-pc sample. *Bottom right:* $(G_{BP} - G_{RP}, G_{RP})$ colour-magnitude magnitude for our 30-pc sample. The "classical" WDs from von Hippel (1998) are shown as black dots. The candidates from Schilbach & Röser (2012) are highlighted with their ID number. The Hyades single and binary WDs from Gaia Collaboration et al. (2018a) are depicted as red and blue circles, respectively.

---

[2]  http://www.astro.umontreal.ca/~bergeron/CoolingModels/, see also Tremblay et al. (2012) and references therein.

[3]  http://people.sissa.it/~sbressan/CAF09_V1.2S_M36_LT/, see also Tang et al. (2014)

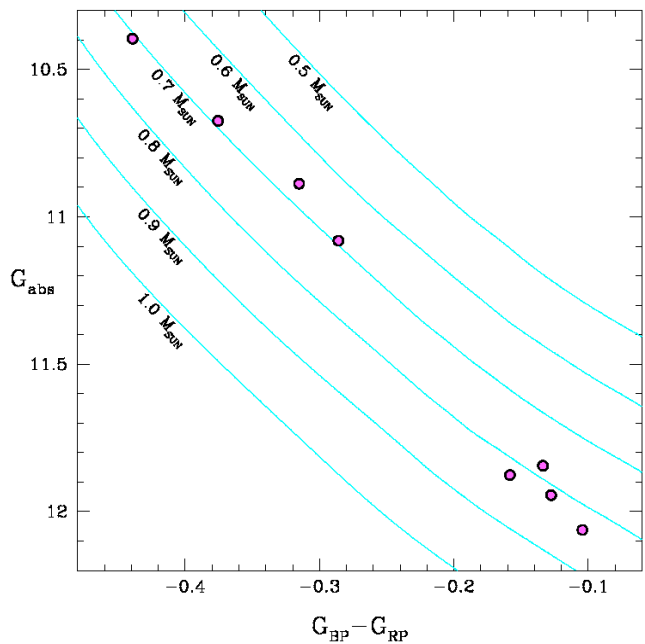

**Fig. 7.** Absolute magnitude of the Hyades white dwarfs as a function of the $G_{BP} - G_{RP}$ colour. The error bars, smaller than the symbols, are always lower than 0.011 and 0.008 in absolute magnitude and $G_{BP} - G_{RP}$ colour respectively. The WD cooling tracks of Bergeron et al. (2011) for various masses are shown in light blue (see text for more details).

The results of our computations are summarised in Table 6. The largest source of error comes from the IFMR: the uncertainty on the initial mass implies an uncertainty on the MS evolutionary time. Compared to this uncertainty, the *Gaia* photometric and parallax errors are negligible for these relatively bright stars. When considering all eight single WDs in Table 6, we infer an age of $640^{+67}_{-49}$ Myr for the Hyades cluster.

This age agrees with the early estimate from the cooling age of the WD by De Gennaro et al. (2009), who quoted 648±45 Myr. Those values are also consistent with the mean age of the cluster derived from isochrone fitting with convective overshoot (625±50 Myr; Maeder & Mermilliod 1981; Mermilliod 1981) and the lithium depletion boundary method (650±70 Myr; Lodieu et al. 2018; Martín et al. 2018). However, it is on the lower side of older ages determined from evolutionary models with enhanced convective overshooting (upper limit of 1.2 Gyr; Mazzei & Pigatto 1988) and models with rotation (750±100 Myr; Brandt & Huang 2015).

After the first submission of this paper, a new article by Salaris & Bedin (2018) makes use of the same sample of single DA WDs. Using a recent estimate of the cluster age from the main-sequence turn off, these authors derived the WD progenitor masses and compared the Hyades IFMR with the global IFMR. It is basically the opposite with respect to what we have done.

# 7. The luminosity and mass functions

In this section, we present the luminosity function as a function of radius from the cluster center and derive the associated mass function including all stellar and sub-stellar members.

## 7.1. Luminosity function

We derive the system luminosity function from our sample of Hyades candidates identified using *Gaia* DR2 data. This sample contains a total of 85, 381, 568, and 710 objects in 3.1, 9, 18, and 30 pc from the cluster center, respectively. We did not attempt to correct the system luminosity function for binaries and postpone this analysis to later *Gaia* releases where astrometric parameters of multiple systems will be incorporated. The impact of binaries is not expected to be negligible as 20–40% of candidates are predicted to be in multiple systems (Duchêne et al. 2013; Gunn et al. 1988; Reid & Mahoney 2000; Reino et al. 2018; Röser et al. 2011). The multiplicity varies as a function of mass, with high-mass stars being more likely in multiple systems. However, the precise impact on the luminosity function is difficult to estimate because multiple systems should be investigated over a wide range of separation and mass ratios.

In Fig. 8, we display the system luminosity function, i.e. the number of objects per absolute magnitude bins ($M_G$), with bin width of one mag scaled to a volume of one cubic parsec for four different annuli from the cluster center. We choose to display the regions within 3.1 pc (core radius), 3.1 to 9 pc (tidal radius), 9 to 18 pc (halo), and 18 to 30 pc. We do not apply any correction to the luminosity function or plot the error bars. However, we note that two of the 10 brightest known pre-*Gaia* members are not recovered in our 30-pc sample for the reasons discussed in Sect. 5.2.4. The LF is also affected to some level by incompleteness at the faint end of *Gaia* because a member at 10 pc will be about one magnitude fainter than a member at the center of the cluster. However, the *Gaia* sample is mainly complete down to the hydrogen-burning limit with an uncertainty of ±0.01 $M_\odot$ because we showed that all brown dwarfs with indices are not catalogued in *Gaia* DR2 (Tables 1–2). Models predict that a 0.072 $M_\odot$ Hyades member has $G \sim 19.65$ mag, making it detectable up to the tidal radius. Further discussion on the incompleteness is provided in the next section (Sect. 7.2). However, the completeness of *Gaia* DR2 is a function of magnitude, position on the sky density (Section 3 and Figure 3 in Arenou et al. 2018). As stated in Sect. 5.1 we limit our study to candidates within 30 pc because the tidal radius may be higher (for example as we have not counted binaries, dust, etc..) and to look for objects that have been stripped off but continue to have the Hyades kinematics. Therefore, from the distribution of sources that passed the membership criteria, we counted the numbers of objects in three annuli (40–50 pc, 50–60 pc, and 60–70 pc) and divided by the volume, yielding numbers of contaminants in the range $3.84$–$7.59 \times 10^{-4}$ per cubic parsec. Therefore, we predict 0.15–0.25, 1.0–2.2, 8.2–16.3, and 34.0–67.3 contaminants among the 85, 381, 568, and 710 sources in the 3.1, 3.1–9, 9–18, and 18–30 pc volumes, respectively

We should add 10 brown dwarfs to our luminosity function but we do not have *Gaia* magnitudes because they are too faint to be detected (Table 2; Sect. 4). We estimated their magnitudes with the equation in Sect. 4.1 (Smart et al. 2017b). We find 1, 1, 4, 1, and 1 objects in the 16–17, 18–19, 19–20, 21–22, and 23–24 magnitude bins, respectively (red lines in Fig. 8), scaled to the volume and distance from the cluster center. We note that there is only one brown dwarf in the 18–30 pc annulus, all the others being within 14 pc.

Among the WDs, we have two in the 3.1 pc core annulus (EGGR 36, EGGR 37) and another nine within the tidal radius (EGGR 29, HZ 14, LP 474-185, HG7-85, EGGR 316, HZ 7, HD 27483), including two known binaries V 471 Tau and HZ 9. Two other WDs (HZ 4 and GD 52) lie between 14.9 and 16.0 pc.

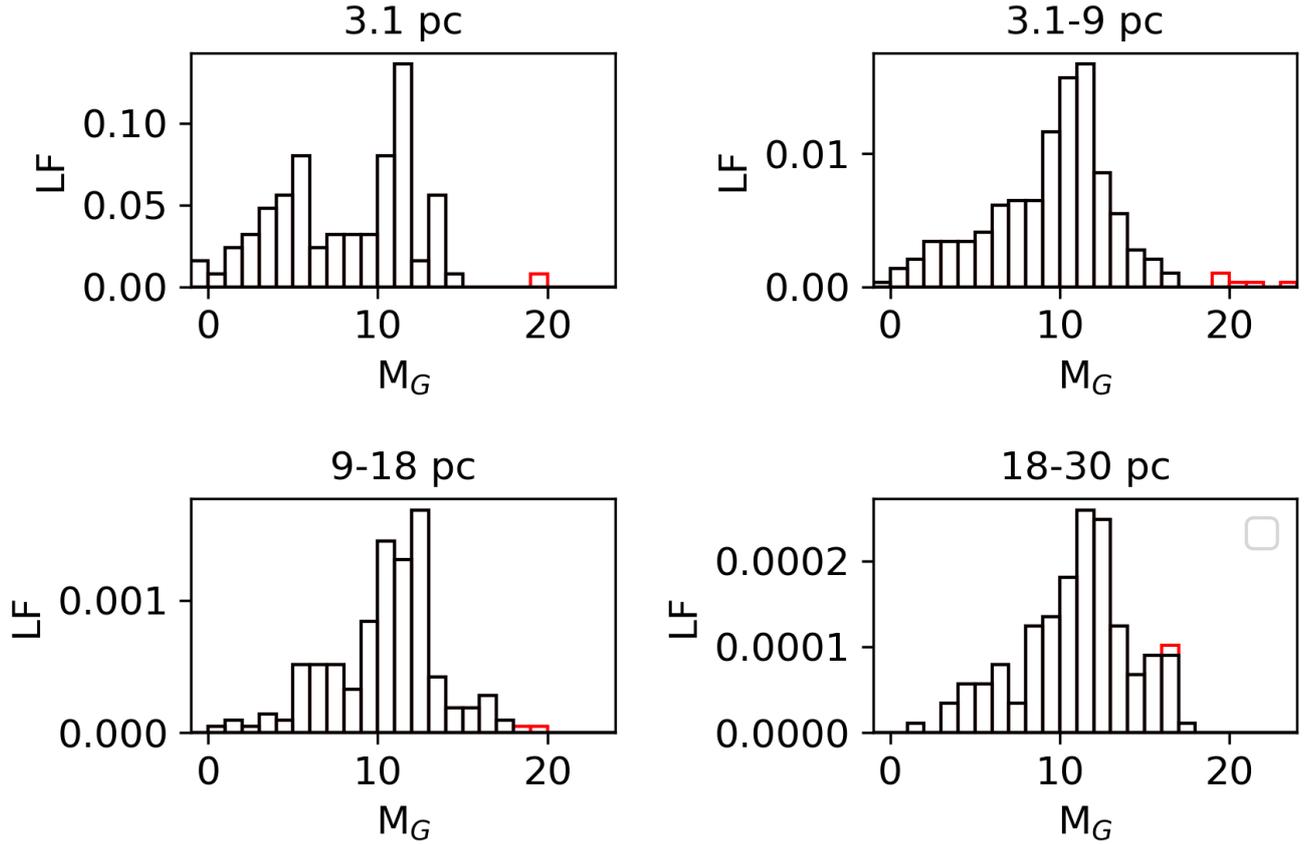

**Fig. 8.** Luminosity functions for radius intervals of 0–3.1 pc, 3.1–9 pc, 9-18 pc, and 18-30 pc from the cluster center scaled to a volume of one cubic parsec. We added the brown dwarfs undetected in *Gaia* DR2 in red.

The remaining two WDs (GD 77 and G 74-15B) are 31–32 pc away from the cluster center, which we reject as members of the cluster. None of the WDs have *Gaia* radial velocity.

We observe distinct shapes of the Hyades luminosity function in different regions of the cluster, from the core to the tidal radius, and the halo as previously reported by studies using Hipparcos and ground-based data. In the central 3.1 pc, we observe two peaks at $M_G = 5.0$–$6.0$ mag and $11$–$12$ mag, corresponding to solar-type stars ($0.9\pm0.1$ $M_\odot$) and M2–M4 dwarfs ($\sim0.2$–$0.3$ $M_\odot$) before it decreases sharply with only one brown dwarf below $M_G = 16$ mag (HyaL5). The luminosity function increases smoothly until it peaks at $M_G \sim 12$ mag within the tidal radius and the brown dwarf bins are populated. Therefore, the lack of brown dwarfs in the core of the Hyades is likely of physical grounds because *Gaia* is most likely complete down to 0.072 $M_\odot$ in magnitude space (with the biases described above). The peak of the luminosity function remains constant around $M_G \sim 12$ mag (M2–M4) beyond the core radius. Assuming magnitude as a proxy for mass we find the relative number per cubic parsec of high-mass stars ($\geq 1.4$ $M_\odot$) in the 3.1–9 pc annulus is 14.75 times lower than in the core. The number of high-mass stars decreases by a factor of $\sim30$ and $\sim10$ in the next two annuli, while the number of solar-type stars (0.7–1.3 $M_\odot$) and low-mass stars (0.1–0.5 $M_\odot$) decrease by a factor of 14 (5.4) and 10.6 (6.3), respectively. The relative density of low-mass stars to high-mass stars is 2.6 in the core radius but increases to 9.2 and 30.2 in the tidal radius and halo, respectively. The luminosity function increases steadily in the 9–18 pc region until it reaches a broader peak at $M_G = 13$ mag, one magnitude fainter than at closer radii, and remains so in the halo.

### 7.2. Mass function

To convert magnitudes into masses, we need a mass-luminosity relation over a wide mass range, from A-type stars down to the sub-stellar regime. To derive the most reliable present-day mass function, we would require eclipsing binaries with accurate masses and radii at the age of the Hyades. This information is not available despite recent advances thanks to Kepler/K2 which identified transiting planets orbiting members of the Hyades (Ciardi et al. 2018; David et al. 2016; Livingston et al. 2018; Mann et al. 2016, 2018). We selected both Hyades planet-host stars (EPIC 247589423 and EPIC 210490365) as kinematic candidates with distances from the cluster center of 12.96 pc and 4.45 pc of the cluster, respectively. We remark that EPIC 247589423 lies beyond the tidal radius of the cluster but has a radial velocity from *Gaia* consistent with the cluster. Since we do not have an empirical calibration, we adopt a model-dependent mass-magnitude relation and note that all the results in this section follow from that. We considered two types of models at an age of 650 Myr to convert observables into masses: the Padova isochrones (PARSEC v1.2S + COLIBRI PR16; Bressan et al. 2012; Marigo et al. 2013, 2008, 2017; Rosenfield et al. 2016)[4] and the BT-Settl models (Allard et al. 2012; Baraffe et al.

---

[4]  http://stev.oapd.inaf.it/cgi-bin/cmd

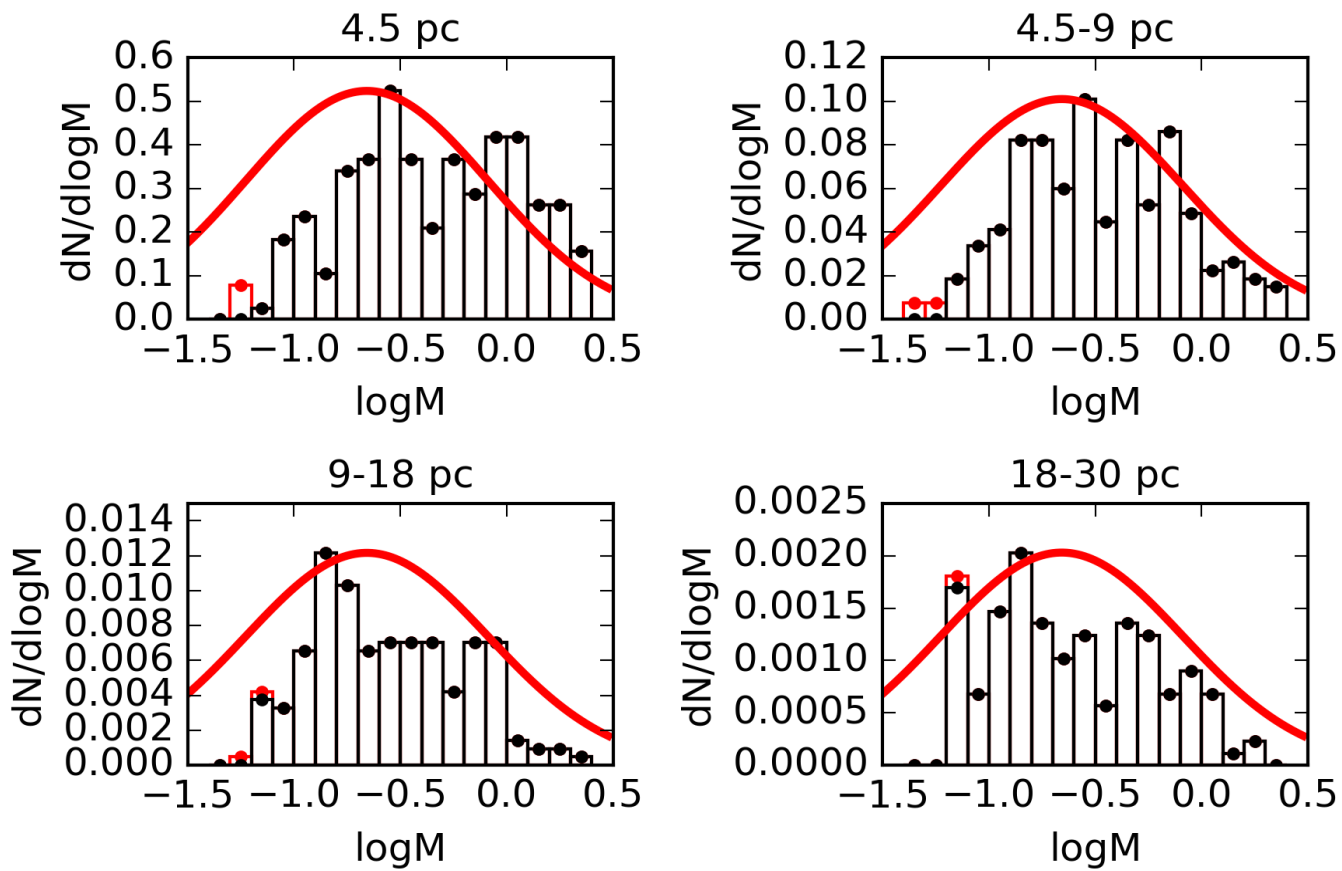

**Fig. 9.** Mass functions for annuli of 0–3.1 pc, 3.1–9 pc, 9–18 pc, and 18–30 pc from the cluster center scaled to a volume of one cubic parsec and mass bins of 0.1 dex in logarithmic units assuming an age of 650 Myr. Overplotted with a thick red line is the lognormal form of the field mass function (Chabrier 2003) normalised to one and multiplied by the most populated bin in the volume considered.

2015)[5]. We also produced combined models for ages of 600 Myr and 700 Myr.

We compared the masses predicted by the BT-Settl isochrones to the dynamical masses of field late-M and L dwarfs (Dupuy & Liu 2017). We should bear in mind the difference in age between Hyades members (650±100 Myr) and field dwarfs (>1 Gyr). Based on the four candidates of Hogan et al. (2008) with optical spectral types recovered in our *Gaia* sample (Hya01, Hya03, Hya06, and Hya08), we would infer masses of 0.098±0.011 and 0.085±0.010 $M_\odot$ for M8–M8.5 and M9–L0.5 dwarfs (Dupuy & Liu 2017) while the models predict masses of 0.077 and 0.065–0.068 $M_\odot$. Considering the difference in ages and the range in dynamical masses, the agreement between observed and model-dependent masses is acceptable to proceed with the derivation of the mass function bearing in mind these caveats.

We opted to merge both models to cover the full range of masses. We kept the Padova and BT-Settl models above and below 1.4 $M_\odot$ ($M_G \sim 3.1$ mag), respectively. We find small differences between both models at 1.4 $M_\odot$: $\log(L/L_\odot) = 0.62$ vs 0.60 dex in luminosities, 6918 K vs 6724 K in effective temperatures, and $\log g = 4.28$ vs 4.25 dex in gravities. We have now a mass-luminosity relation from 2.6 $M_\odot$ down to 0.05 $M_\odot$ equivalent to absolute $G$ magnitudes of $-2.78$ mag and 19.03 mag, respectively. The full magnitude range of apparent *Gaia* magnitudes (3–21 mag) is therefore covered as are brown dwarfs in the Hyades.

We plot the (system) mass function in Fig. 9 counting the number of objects per volume cubic parsec and per bins of 0.1 dex in logarithmic units of mass, assuming an age of 650 Myr. We obtained the masses for the brown dwarfs in a different manner because we have detection of lithium in absorption at 6707.8Å for some of them, placing their mass in the 0.05–0.06 $M_\odot$ range (Baraffe et al. 2015; Basri et al. 1996; Rebolo et al. 1992). We place Hya02 (M8.5) and Hya11 (L3.0) in the 0.06–0.07 $M_\odot$ mass bin because they have depleted their lithium, while the T dwarfs from Bouvier et al. (2008) have most likely masses below 0.05 $M_\odot$. Therefore, we place two, seven, and two sub-stellar members in the 0.06–0.07, 0.05–0.06, and 0.04–0.05 $M_\odot$ intervals, respectively, in addition to all the *Gaia* members.

We plot the Hyades (system) mass function for four annuli (3.1, 3.1–9, 9–18, and 18–30 pc) in Fig. 9. We overplot the lognormal form of the field mass function (red line) from Chabrier (2003), normalised to the value of the most populated bin of objects in the volume under consideration. We observe that the Hyades mass function is not reproduced by the field mass function in any of the regions. In the core of the cluster, we observe an excess of high-mass stars ($\geq$1.4 $M_\odot$) with respect to the field while low-mass stars are under-represented. The mass function within the tidal radius is relatively well fit by the lognormal form of the field IMF, except for stars below 0.1 $M_\odot$ and brown dwarfs. In the halo, the high-mass stars are clearly underrepresented because their number is scarce while low-mass stars start to dominate. The lack of brown dwarfs remain evident in all regions, which we attribute to the mass segregation and the incompleteness of previous ground-based surveys in the sub-stellar regime due to a combination of a lack of sensitivity and limited spatial coverage (Bouvier et al. 2008; Pérez-Garrido et al. 2017, 2018).

As stated above, the system mass function is not corrected for binaries, which we cannot do at this stage. The upcoming release of *Gaia* will contribute to the correction but a wider spectrum of physical separations must be probed before a "resolved" mass function can be derived. To estimate the impact of binaries on the shape of the mass function, we have compared the system and resolved mass functions of the field (Figure 1 and equations 17+18 in Chabrier 2003). Assuming that the multiplicity of Hyades members is comparable to the binary fractions in the solar neighbourhood, we should apply the multiplicative factors of 0.96, 1.0, 1.14, 2.17, and 3.33 to the numbers of members at masses of 2.0, 1.0, 0.5, 0.3, and 0.1 $M_\odot$, respectively.

We infer a total mass of 61 $M_\odot$ in 85 systems within a radius of 3.1 pc from the cluster center assuming an age of 650 Myr and the models described earlier. The total mass is about 218, 292, and 343 $M_\odot$ in 381, 568, and 710 systems within 9, 18, and 30 pc from the cluster center, respectively. We find a slightly larger number of systems within the central 9 pc than Röser et al. (2011) with a smaller total mass (381 vs 364 systems and 218 $M_\odot$ vs 275 $M_\odot$). We find another ~70 $M_\odot$ in the halo (9–18 pc), which is two-thirds of the one reported by Röser et al. (2011) but a similar additional mass budget in the 18–30 pc annulus (~50 $M_\odot$ vs 60 $M_\odot$). The most likely explanation is that the high astrometric precision of the *Gaia* data has cleaned up the contaminants in the halo of the cluster but possibly not fully beyond where we still expect significant contamination (lower right panel in Fig. 8). We caution that those total mass estimates do not take into account multiple systems. Assuming a multiplicity fraction of 20–40% and an average mass of the secondary equal to two-thirds of the primary, the correction factor to apply to the aforementioned numbers would be of order of 16–27%, which is not negligible at all. We inferred a tidal radius of 8.3 pc from *Gaia* DR2 alone, in agreement with Röser et al. (2011). Accounting for 20–40% binaries among Hyades members, the tidal radius of the cluster may increase to 10.5–13.5 pc depending on the binary fraction and mass ratios. We also investigated the impact of the uncertainty on the age of the Hyades by calculating the total mass in each distance annulus for isochrones of 600 Myr and 700 Myr, corresponding to about 10% uncertainty. At those ages, the isochrones are very similar over a wide range of masses, except above 1.8 $M_\odot$ and below the hydrogen-burning limit. Both mass intervals contain a limited number of members. The impact is less than 1% on the total mass, therefore minimal, and much lower than the effect of multiplicity discussed above.

# 8. Discussion: a 3D view of the Hyades

## 8.1. Distribution in space

With the availability of accurate astrometry from *Gaia* DR2, we are now able to draw a 3D map in galactic coordinates for all bona-fide members identified from their kinematics. We depict the distribution of all 1764 member candidates in Fig. 10 but limited the plots to the central regions where we overplotted four circles in cyan representing the core (3.1 pc), tidal (9 pc), halo (18 pc), and 30 pc radius. We added in red in Fig. 10 the galac-

| dist | A | F | G | K | M | L | T |
|------|-----|-----|---------|-----------|------------|--------|------------|
| 3.1  | 8  | 11 | 11 (44) | 12 (160)  | 40 (864)   | 1 (17) | 0 ($\geq$121) |
| 9.0  | 25 | 29 | 29 (112)| 75 (406)  | 209 (2277) | 5 (44) | 2 ($\geq$319) |
| 18.0 | 29 | 31 | 42 (124)| 110 (450) | 330 (2434) | 7 (47) | 2 ($\geq$341) |
| 30.0 | 31 | 36 | 51 (144)| 133 (522) | 419 (2826) | 8 (54) | 2 ($\geq$396) |

tic coordinates of 10 brown dwarfs discussed earlier for which we have ground-based parallaxes, except in the case of the two mid-L candidates identified by Schneider et al. (2017) where we included their photometric distances for completeness. These are the only two sources without parallaxes in our sample in addition to Hya04 which has no parallax in *Gaia* DR2.

The (0,0,0) in galactic coordinates represent the position of the Sun and (1,0,0) is a unit vector pointing to the galactic centre. We observe that the cluster shows a centrally-concentrated group of stars with two tails in the X and Y directions (Fig. 10). The extension of the cluster in Y,Z is consistent with the extension found in the Hipparcos data (Perryman et al. 1998). We refer to that work for a detailed interpretation of the possible causes for the spatial distribution of cluster members. The elongation of the cluster in the X direction is increasing dramatically from 3.1 to 9 pc from the cluster center with the population in the tail to the Galactic anti-center increasing quickly between 3.1, 9, and 18 pc. The overall extension of about 80 pc, from −20 pc to −100 pc is also consistent with the results of Hipparcos (Perryman et al. 1998) but is now homogeneously distributed with the presence of a tail towards negative values of $b_x$. We note that the cluster appears also elongated in the Y direction beyond 18 pc, in the region of the halo of the cluster where we might identify a mix of members of the cluster and the Hyades moving group (Boss 1908; Eggen 1958; Famaey et al. 2007; Zuckerman & Song 2004).

In Fig. 11, we plot the 3D distribution of galactic velocities ($b_v$ in km s$^{-1}$) of the 192 Hyades members at distances less than 30 pc from the cluster center and *Gaia* radial velocities ($G = 5.6$–13.6 mag). The size of the sample corresponds approximately to the size of the Hipparcos sample but with more accurate radial velocities. We observe an extension along the X-axis with a small sub-group of objects towards negative $v_x$. We confirm the conclusions drawn in Perryman et al. (1998).

## 8.2. Mass segregation

The differences seen in the luminosity function in different annuli from the cluster center clearly indicate a mass segregation in the Hyades (Fig. 8), in fact extensively discussed in the literature (e.g. Bouvier et al. 2008; Perryman et al. 1998; Röser et al. 2011) and reproduced through numerical simulations by the loss of the lowest mass members over a timescale of a few 100 Myr (de La Fuente Marcos 1995; Kroupa 1995; Terlevich 1987).

In Fig. 12 we display the distance from the center of the cluster as a function of the *Gaia* magnitude $G$ for the 3.1, 9, 18, and 30 pc radii from the cluster center. We overplot the Hyades brown dwarfs in red. In these diagrams, as we can assume the objects are of the similar age, composition and distance then magnitude will act as a proxy for mass. We can clearly see the equal number of stars at different masses within the central ∼3 pc whereas low-mass stars become more numerous beyond 3 pc. At distances larger than 6–7 pc, the number of high-mass stars become negligible (Fig. 12). We observe a possible bias of brown dwarf members on the side towards us, suggesting that we are more complete nearby than the other side, which can be a consequence of the depth of photometric surveys. These plots represent another proof of the mass segregation present in the Hyades. We observe that one of the known brown dwarfs lies at 2.6 pc from the cluster (red dots in Fig. 10). Most of them lie within the tidal radius and the furthest at about 20 pc. Their distribution seems to match the distribution of M-type members.

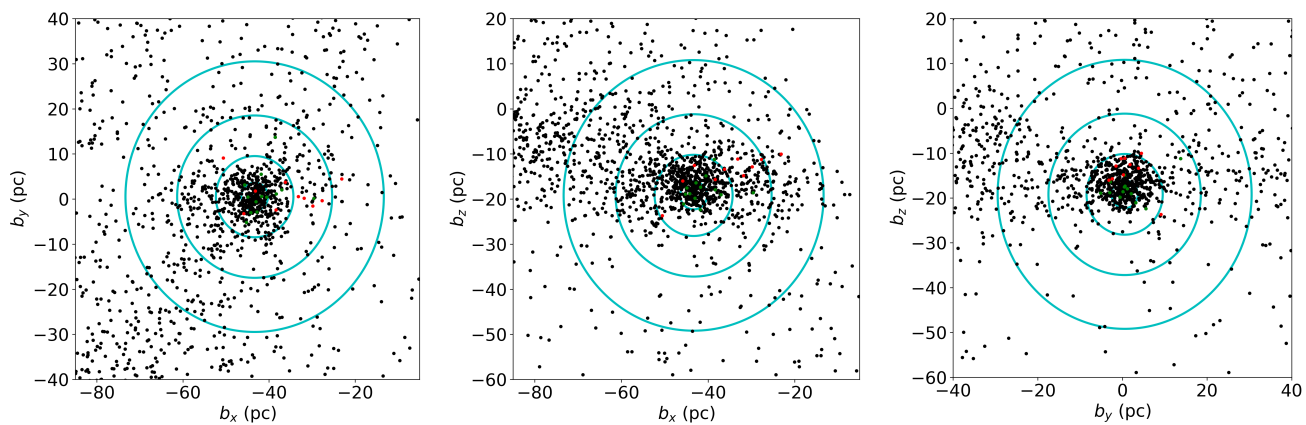

**Fig. 10.** 3D coordinates in space ($\mathbf{b}_c$ in pc) of the 1764 Hyades stellar (black) and 10 sub-stellar (red) members with four different annuli from the cluster center drawn in cyan: 3.1, 9, 18, and 30 pc. White dwarfs are highlighted in green.

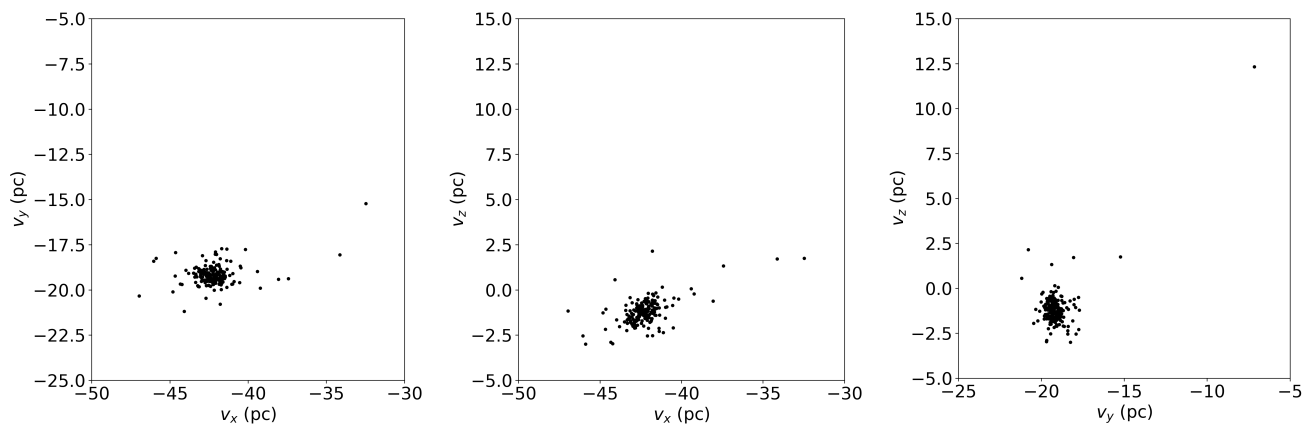

**Fig. 11.** 3D velocities of 192 Hyades members within 30 pc with *Gaia* RVs. One object lies outside the limits of the ($v_x, v_y$) plot with $v_x = -17.26$ pc (the outlier at the top-right on the right-hand side diagram).

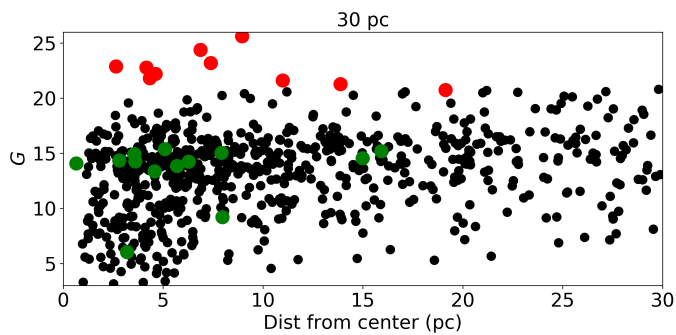

**Fig. 12.** Distance from the cluster center in 3D space vs $G$ magnitude for Hyades members selected kinematically up to a radius of 30 pc. We overplotted brown dwarfs with ground-based parallaxes and white dwarfs as red and green dots, respectively.

To illustrate the effect of segregation in a more quantitative manner, we compute the ratio $R$ of stars to brown dwarfs as defined by Andersen et al. (2008): the sum of all objects in the interval 0.08–1.0 $M_\odot$ is divided by the number of brown dwarfs (0.03–0.08 $M_\odot$). We emphasise that the bin of brown dwarfs is yet incomplete in the Hyades because the coolest members are early-T dwarfs with model-dependent masses of the order of 0.05 $M_\odot$ (Bouvier et al. 2008) and confirmed L-type brown dwarf with lithium in absorption at 6707.8Å and masses in the range 0.05–0.06 $M_\odot$ (Lodieu et al. 2018; Martín et al. 2018). The 0.03–0.05 $M_\odot$ bin remains currently unexplored in the Hyades cluster; thus all number are upper limits. As described in Sect. 4 we have confirmed spectroscopically, as well as through ground-based parallaxes, the presence of eight brown dwarfs in the cluster to which we should add the two L5+L6 dwarfs from Schneider et al. (2017). Among these 10 brown dwarfs, three are located in the core, four in the 3.1–9 pc region, two in the 9–18 pc annulus, and one even farther. We will take into account this fact in the following computations.

In the central 3.1 pc, we find an upper limit of R = 62 with the lack of object with a model-dependent mass below 0.08 $M_\odot$ to which we added the only brown dwarf HyaL5. This ratio decreases with larger radii, going from R = 22.1 to 18.4, and 14.6 at 9, 18, and 30 pc from the cluster center, respectively (after correction for the numbers of brown dwarfs). However, we argue that the search for sub-stellar members is incomplete beyond the tidal radius either because previous surveys ignored those regions or were not sensitive enough. Therefore, the last two ratios are clearly upper limit rather than exact values. Nonetheless, the evolution of the ratio as a function of distance from the cluster center is another proof of the mass segregation in the Hyades cluster even though we shall keep in mind that these values are upper limits. Those ratios should be compared to the values of 3.3–8.5 and 4.9 derived for star-forming regions (age = 1–2 Myr) and the Pleiades (120 Myr), respectively (Andersen et al. 2008), demonstrating the loss or incompleteness of sub-stellar members in the central part of the Hyades

We extrapolated the current sample of stars within 8 pc of the Sun to investigate the numbers of solar-type, low-mass, and brown dwarfs that may have been lost due to dynamical evolution. This sample is well reproduced by field mass functions (Chabrier 2003; Kroupa et al. 2013). From Fig. 11 of Kirkpatrick et al. (2012), the compilation of all known members within 8 pc of the Sun gives 11 WDs, 4 A, 2 F, 8 G, 29 K, 157 M, and 3 L dwarfs with a lower numbers of 22 T dwarfs. We assume that all F-type members of the Hyades have been identified because their lifetime is larger than the age of the cluster, which might not be the case for all A stars. However, if we consider that all A-type stars have been identified, the extrapolations of the numbers of GKMLT dwarfs would be decreased by a factor of two approximately because we find roughly the same numbers of A and F stars. The same remark would apply if we consider the G stars because we find about the same numbers of F and G dwarfs in all annuli, implying that our extrapolations would decrease by a factor of approximately four. We note that 71 Tauri (F0V) is missing from our final catalogue (Sect. 5.2.4), hence affecting our estimates by at most ∼5%. The other bright object missing ($\Theta^1$ Tauri) is a giant, hence, does not affect our estimates. These differences could be due to several effects, among which mass segregation or stellar formation because field stars have an average age of 2–4 Gyr several times the age of the Hyades during which the star formation may have changed.

We list the numbers of observed dwarfs as a function of spectral type and distance from the cluster center in Table 7 and give the extrapolated numbers in parenthesis, assuming the same conditions. We observe that the cluster has retained about 25% and 18% of its G and K dwarfs within its tidal radius and may have lost 90% of its M dwarfs. Numerical simulations of the dynamical evolution of clusters by Adams & Myers (2001) suggest that a 650 Myr-old cluster may have lost ∼65% and 70–90% of its initial stellar and sub-stellar populations, fairly consistent with our analysis. The numbers of L and T dwarfs in the Hyades are quite low, which may be due to dynamical evolution but also to the lack of sensitivity and limited areal searches of previous studies (e.g. Bouvier et al. 2008; Hogan et al. 2008). We find five L dwarfs and two T dwarfs within the tidal radius, while we should have 44 and more than 319 based on the sample of L and T dwarfs within 8 pc of the Sun. The upcoming Large Synoptic Survey telescope (LSST) might shed some light on this issue (Ivezic et al. 2008).

# 9. Conclusions

We presented updated parameters for the Hyades cluster as well as a revised census of member candidates up to 30 pc from the cluster center in the 2.50–0.04 $M_\odot$ mass interval. The sample within the tidal radius of the cluster is 99% complete while the 18-pc and 30-pc samples might suffer from a level of contamination of ≤ 3% and ∼5-10%, respectively. We combined the *Gaia* DR2 astrometry with ground-based parallaxes from the Liverpool telescope to produce the first 3D map of the cluster[6].

We summarise the main results of our work:

- we derive a mean distance of 47.03±0.20 pc for the cluster and a mean velocity of 46.38±0.12 km s$^{-1}$ from members within the tidal radius

- we present ground-based parallaxes derived with the IO:I infrared camera on the 2-m robotic Liverpool telescope for eight of the 10 known Hyades brown dwarfs.

- we identified a total of 85, 381, 568, and 710 astrometric members within 3.1 (core radius), 9 (tidal radius), 18 (halo), and 30 pc from the center of the Hyades with the implementation of the kinematic method described by Perryman et al. (1998). This sequence of Hyades members at 650 Myr constitutes a benchmark in colour-magnitude diagrams involving optical to infrared magnitudes.

- we confirm the membership of previously-known white dwarfs and assess the membership of other possible pre-DR2 white dwarf candidates. We derive an age of $640^{+67}_{-49}$ Myr from the nine single white dwarfs by comparing their *Gaia* photometry with state-of-the-art models, in agreement with the age derived with the lithium depletion boundary method.

- we derive the luminosity and mass functions of the Hyades in the $G = 3$–26 mag range translating into masses in the 2.5–0.04 $M_\odot$ interval. The shapes vary as a function of the distances from the center.

- we derive 3D positions for all stellar and sub-stellar members and show a 3D map of the Hyades. The cluster center is identified by −43.83±0.18, +0.42±0.11,−17.05±0.09 pc in $b_x$, $b_y$, and $b_z$, respectively.

- we observe a spatially-concentrated distribution of stellar members in 3D space with the extension of members towards the direction of the Galactic center in velocity space and along the $(b_x,b_y)$ axis in coordinate space, as previously reported in studies exploiting Hipparcos and ground-based data.

- we find that the cluster has clearly suffered mass segregation with the low-mass members being on average further away from the center than high-mass and solar-type members.

Our study has ignored multiplicity over the full mass range. The next *Gaia* release should include information on binaries with some preliminary orbits to study the multiplicity s a function of mass and improve the mass determinations of the Hyades members to derive a more accurate mass function.

Finally, spectroscopic follow-up at medium-resolution is required to infer radial velocities of *Gaia* member candidates fainter than $G \sim 13.5$ mag and clean the astrometric sequence beyond the tidal radius.

*Acknowledgements.* NL and APG were financially supported by the Spanish Ministry of Economy and Competitiveness (MINECO) under the grants AYA2015-69350-C3-2-P and AYA2015-69350-C3-3-P. The authors would like to thank Jonathan Gagné and Jos de Bruijne for useful discussions during this work as well as Pierre Bergeron for having made available, before publication, the WD cooling tracks complete with *Gaia* magnitudes. This research has made use of the Simbad and Vizier databases, operated at the center de Données Astronomiques de Strasbourg (CDS), and of NASA's Astrophysics Data System Bibliographic Services (ADS). This research has also made use of some of the tools developed as part of the Virtual Observatory. This work has made use of data from the European Space Agency (ESA) mission *Gaia* (https://www.cosmos.esa.int/gaia), processed by the *Gaia* Data Processing and Analysis Consortium (DPAC,

---

[6] A movie is available at https://www.dropbox.com/s/19m4q08eenkzq8z/animati...

https://www.cosmos.esa.int/web/gaia/dpac/consortium). Funding for the DPAC has been provided by national institutions participating in the Gaia Multilateral Agreement and in particular the support of the Agenzia Spaziale Italiana under contract I/058/10/0.

The Liverpool Telescope is operated on the island of La Palma by Liverpool John Moores University in the Spanish Observatorio del Roque de los Muchachos of the Instituto de Astrofisica de Canarias with financial support from the UK Science and Technology Facilities Council. Funding for the Sloan Digital Sky Survey IV has been provided by the Alfred P. Sloan Foundation, the U.S. Department of Energy Office of Science, and the Participating Institutions. SDSS-IV acknowledges support and resources from the Center for High-Performance Computing at the University of Utah. The SDSS web site is www.sdss.org. SDSS-IV is managed by the Astrophysical Research Consortium for the Participating Institutions of the SDSS Collaboration including the Brazilian Participation Group, the Carnegie Institution for Science, Carnegie Mellon University, the Chilean Participation Group, the French Participation Group, Harvard-Smithsonian Center for Astrophysics, Instituto de Astrofísica de Canarias, The Johns Hopkins University, Kavli Institute for the Physics and Mathematics of the Universe (IPMU) / University of Tokyo, Lawrence Berkeley National Laboratory, Leibniz Institut für Astrophysik Potsdam (AIP), Max-Planck-Institut für Astronomie (MPIA Heidelberg), Max-Planck-Institut für Astrophysik (MPA Garching), Max-Planck-Institut für Extraterrestrische Physik (MPE), National Astronomical Observatories of China, New Mexico State University, New York University, University of Notre Dame, Observatário Nacional / MCTI, The Ohio State University, Pennsylvania State University, Shanghai Astronomical Observatory, United Kingdom Participation Group, Universidad Nacional Autónoma de México, University of Arizona, University of Colorado Boulder, University of Oxford, University of Portsmouth, University of Utah, University of Virginia, University of Washington, University of Wisconsin, Vanderbilt University, and Yale University. This publication makes use of data products from the Two Micron All Sky Survey, which is a joint project of the University of Massachusetts and the Infrared Processing and Analysis Center/California Institute of Technology, funded by the National Aeronautics and Space Administration and the National Science Foundation.

The UKIDSS project is defined in Lawrence et al. (2007). UKIDSS uses the UKIRT Wide Field Camera Casali et al. (WFCAM; 2007). The photometric system is described in Hewett et al. (2006), and the calibration is described in Hodgkin et al. (2009). The pipeline processing and science archive are described in Irwin et al. (2009, in prep) and Hambly et al. (2008).

This publication makes use of data products from the Wide-field Infrared Survey Explorer, which is a joint project of the University of California, Los Angeles, and the Jet Propulsion Laboratory/California Institute of Technology, and NEOWISE, which is a project of the Jet Propulsion Laboratory/California Institute of Technology. WISE and NEOWISE are funded by the National Aeronautics and Space Administration.

The Pan-STARRS1 Surveys (PS1) and the PS1 public science archive have been made possible through contributions by the Institute for Astronomy, the University of Hawaii, the Pan-STARRS Project Office, the Max-Planck Society and its participating institutes, the Max Planck Institute for Astronomy, Heidelberg and the Max Planck Institute for Extraterrestrial Physics, Garching, The Johns Hopkins University, Durham University, the University of Edinburgh, the Queen's University Belfast, the Harvard-Smithsonian Center for Astrophysics, the Las Cumbres Observatory Global Telescope Network Incorporated, the National Central University of Taiwan, the Space Telescope Science Institute, the National Aeronautics and Space Administration under Grant No. NNX08AR22G issued through the Planetary Science Division of the NASA Science Mission Directorate, the National Science Foundation Grant No. AST-1238877, the University of Maryland, Eotvos Lorand University (ELTE), the Los Alamos National Laboratory, and the Gordon and Betty Moore Foundation.

# References

Abolfathi, B., Aguado, D. S., Aguilar, G., et al. 2018, ApJS, 235, 42
Adams, F. C. & Myers, P. C. 2001, ApJ, 553, 744
Allard, F., Homeier, D., & Freytag, B. 2012, Royal Society of London Philosophical Transactions Series A, 370, 2765
Andersen, M., Meyer, M. R., Greissl, J., & Aversa, A. 2008, ApJL, 683, L183
Arenou, F., Luri, X., Babusiaux, C., et al. 2018, A&A, 616, A17
Artigau, É., Gagné, J., Faherty, J., et al. 2015, ApJ, 806, 254
Baraffe, I., Homeier, D., Allard, F., & Chabrier, G. 2015, A&A, 577, A42
Barnsley, R. M., Jermak, H. E., Steele, I. A., et al. 2016, Journal of Astronomical Telescopes, Instruments, and Systems, 2, 015002
Barrado y Navascués, D., Bouvier, J., Stauffer, J. R., Lodieu, N., & McCaughrean, M. J. 2002, A&A, 395, 813
Basri, G., Marcy, G. W., & Graham, J. R. 1996, ApJ, 458, 600
Bergeron, P., Wesemael, F., Dufour, P., et al. 2011, ApJ, 737, 28
Boesgaard, A. M. & Friel, E. D. 1990, ApJ, 351, 467
Boss, L. J. 1908, AJ, 26, 31
Boucher, A., Lafrenière, D., Gagné, J., et al. 2016, ApJ, 832, 50
Bouvier, J., Kendall, T., Meeus, G., et al. 2008, A&A, 481, 661
Bouvier, J., Stauffer, J. R., Martín, E. L., et al. 1998, A&A, 336, 490
Brandt, T. D. & Huang, C. X. 2015, ApJ, 807, 58
Bressan, A., Marigo, P., Girardi, L., et al. 2012, MNRAS, 427, 127
Bryja, C., Humphreys, R. M., & Jones, T. J. 1994, AJ, 107, 246
Casali, M., Adamson, A., Alves de Oliveira, C., et al. 2007, A&A, 467, 777
Casewell, S. L., Littlefair, S. P., Burleigh, M. R., & Roy, M. 2014, ArXiv e-prints
Cayrel de Strobel, G., Crifo, F., & Lebreton, Y. 1997, in ESA Special Publication, Vol. 402, Hipparcos - Venice '97, ed. R. M. Bonnet, E. Høg, P. L. Bernacca, L. Emiliani, A. Blaauw, C. Turon, J. Kovalevsky, L. Lindegren, H. Hassan, M. Bouffard, B. Strim, D. Heger, M. A. C. Perryman, & L. Woltjer, 687–688
Chabrier, G. 2003, PASP, 115, 763
Chambers, K. C., Magnier, E. A., Metcalfe, N., & 103 co-authors. 2016, ApJ
Ciardi, D. R., Crossfield, I. J. M., Feinstein, A. D., et al. 2018, AJ, 155, 10
Cutri, R. M. & et al. 2014, VizieR Online Data Catalog, 2328, 0
Cutri, R. M., Skrutskie, M. F., van Dyk, S., et al. 2003, 2MASS All Sky Catalog of point sources, 2246
David, T. J., Conroy, K. E., Hillenbrand, L. A., et al. 2016, AJ, 151, 112
de Bruijne, J. H. J., Hoogerwerf, R., & de Zeeuw, P. T. 2001, A&A, 367, 111
De Gennaro, S., von Hippel, T., Jefferys, W. H., et al. 2009, ApJ, 696, 12
de La Fuente Marcos, R. 1995, A&A, 301, 407
Dias, W. S., Alessi, B. S., Moitinho, A., & Lépine, J. R. D. 2002, A&A, 389, 871
Dias, W. S., Assafin, M., Flório, V., Alessi, B. S., & Líbero, V. 2006, A&A, 446, 949
Dias, W. S., Monteiro, H., & Assafin, M. 2018, MNRAS
Dias, W. S., Monteiro, H., Caetano, T. C., et al. 2014, A&A, 564, A79
Dobbie, P. D., Kenyon, F., Jameson, R. F., et al. 2002, MNRAS, 329, 543
Duchêne, G., Bouvier, J., Moraux, E., et al. 2013, A&A, 555, A137
Dupuy, T. J. & Liu, M. C. 2017, ApJS, 231, 15
Eggen, O. J. 1958, MNRAS, 118, 65
Eggen, O. J. 1998, AJ, 116, 284
El-Badry, K., Rix, H.-W., & Weisz, D. R. 2018, ApJL, 860, L17
Famaey, B., Pont, F., Luri, X., et al. 2007, A&A, 461, 957
Gagné, J., Faherty, J. K., Cruz, K. L., et al. 2015a, ApJS, 219, 33
Gagné, J., Faherty, J. K., Mamajek, E. E., et al. 2017, ApJS, 228, 18
Gagné, J., Lafrenière, D., Doyon, R., Malo, L., & Artigau, É. 2014, ApJ, 783, 121
Gagné, J., Lafrenière, D., Doyon, R., Malo, L., & Artigau, É. 2015b, ApJ, 798, 73
Gagné, J., Mamajek, E. E., Malo, L., et al. 2018a, ApJ, 856, 23
Gagné, J., Roy-Loubier, O., Faherty, J. K., Doyon, R., & Malo, L. 2018b, ApJ, 860, 43
Gaia Collaboration, Babusiaux, C., van Leeuwen, F., et al. 2018a, A&A, 616, A10
Gaia Collaboration, Brown, A. G. A., Vallenari, A., et al. 2018b, ArXiv e-prints
Gaia Collaboration, Prusti, T., de Bruijne, J. H. J., et al. 2016, A&A, 595, A1
Gaia Collaboration, van Leeuwen, F., Vallenari, A., et al. 2017, A&A, 601, A19
Gebran, M., Vick, M., Monier, R., & Fossati, L. 2010, A&A, 523, A71
Gizis, J. E., Reid, I. N., & Monet, D. G. 1999, AJ, 118, 997
Goldman, B., Röser, S., Schilbach, E., et al. 2013, A&A, 559, A43
Grenon, M. 2000, in IAU Joint Discussion, Vol. 13, IAU Joint Discussion
Gunn, J. E., Griffin, R. F., Griffin, R. E. M., & Zimmerman, B. A. 1988, AJ, 96, 198
Hambly, N. C., Collins, R. S., Cross, N. J. G., et al. 2008, MNRAS, 384, 637
Hanson, R. B. 1975, AJ, 80, 379
Hewett, P. C., Warren, S. J., Leggett, S. K., & Hodgkin, S. T. 2006, MNRAS, 367, 454
Hodgkin, S. T., Irwin, M. J., Hewett, P. C., & Warren, S. J. 2009, MNRAS, 394, 675
Høg, E., Fabricius, C., Makarov, V. V., et al. 2000, A&A, 355, L27
Hogan, E., Jameson, R. F., Casewell, S. L., Osbourne, S. L., & Hambly, N. C. 2008, MNRAS, 388, 495
Ivezic, Z., Axelrod, T., Brandt, W. N., et al. 2008, Serbian Astronomical Journal, 176, 1
Jones, D. H. P. 1971, MNRAS, 152, 231
Kaiser, N., Aussel, H., Burke, B. E., et al. 2002, in Society of Photo-Optical Instrumentation Engineers (SPIE) Conference Series, Vol. 4836, Survey and Other Telescope Technologies and Discoveries, ed. J. A. Tyson & S. Wolff, 154–164
Keenan, P. C. & McNeil, R. C. 1989, ApJS, 71, 245
Kirkpatrick, J. D., Gelino, C. R., Cushing, M. C., et al. 2012, ApJ, 753, 156
Kroupa, P. 1995, MNRAS, 277, 1522
Kroupa, P., Weidner, C., Pflamm-Altenburg, J., et al. 2013, The Stellar and Sub-Stellar Initial Mass Function of Simple and Composite Populations, ed. T. D. Oswalt & G. Gilmore, 115
Lada, C. J. & Lada, E. A. 2003, ARA&A, 41, 57

Lawrence, A., Warren, S. J., Almaini, O., et al. 2007, MNRAS, 379, 1599
Lebreton, Y., Fernandes, J., & Lejeune, T. 2001, A&A, 374, 540
Leggett, S. K., Harris, H. C., & Dahn, C. C. 1994, AJ, 108, 944
Leggett, S. K. & Hawkins, M. R. S. 1989, MNRAS, 238, 145
Lindegren, L., Hernández, J., Bombrun, A., et al. 2018, A&A, 616, A2
Livingston, J. H., Dai, F., Hirano, T., et al. 2018, AJ, 155, 115
Lodieu, N., Boudreault, S., & Béjar, V. J. S. 2014a, MNRAS, 445, 3908
Lodieu, N., McCaughrean, M. J., Barrado Y Navascués, D., Bouvier, J., & Stauffer, J. R. 2005, A&A, 436, 853
Lodieu, N., Pérez-Garrido, A., Béjar, V. J. S., et al. 2014b, A&A, 569, A120
Lodieu, N., Rebolo, R., & Pérez-Garrido, A. 2018, A&A, 615, L12
Luri, X., Brown, A. G. A., Sarro, L. M., et al. 2018, A&A, 616, A9
Maeder, A. & Mermilliod, J. C. 1981, A&A, 93, 136
Malo, L., Artigau, É., Doyon, R., et al. 2014a, ApJ, 788, 81
Malo, L., Doyon, R., Feiden, G. A., et al. 2014b, ApJ, 792, 37
Malo, L., Doyon, R., Lafrenière, D., et al. 2013, ApJ, 762, 88
Mann, A. W., Gaidos, E., Mace, G. N., et al. 2016, ApJ, 818, 46
Mann, A. W., Vanderburg, A., Rizzuto, A. C., et al. 2018, AJ, 155, 4
Marigo, P., Bressan, A., Nanni, A., Girardi, L., & Pumo, M. L. 2013, MNRAS, 434, 488
Marigo, P., Girardi, L., Bressan, A., et al. 2008, A&A, 482, 883
Marigo, P., Girardi, L., Bressan, A., et al. 2017, ApJ, 835, 77
Martín, E. L., Lodieu, N., Pavlenko, Y., & Béjar, V. J. S. 2018, ApJ, 856, 40
Mazzei, P. & Pigatto, L. 1988, A&A, 193, 148
Melnikov, S. & Eislöffel, J. 2018, A&A, 611, A34
Mermilliod, J. C. 1981, A&A, 97, 235
Mermilliod, J.-C., Mayor, M., & Udry, S. 2009, A&A, 498, 949
Moraux, E., Bouvier, J., & Stauffer, J. R. 2001, A&A, 367, 211
Newton, E. R., Charbonneau, D., Irwin, J., & Mann, A. W. 2015, ApJ, 800, 85
Pérez-Garrido, A., Lodieu, N., & Rebolo, R. 2017, A&A, 599, A78
Pérez-Garrido, A., Lodieu, N., Rebolo, R., & Chinchilla, P. 2018, A&A, 620, A130
Perryman, M. A. C., Brown, A. G. A., Lebreton, Y., et al. 1998, A&A, 331, 81
Perryman, M. A. C., Lindegren, L., Kovalevsky, J., Hoeg, E., & 15 co-authors. 1997, A&A, 323, L49
Rebolo, R., Martín, E. L., & Magazzù, A. 1992, ApJL, 389, L83
Reid, I. N. & Gizis, J. E. 1997, AJ, 114, 1992
Reid, I. N. & Hawley, S. L. 1999, AJ, 117, 343
Reid, I. N., Kirkpatrick, J. D., Liebert, J., et al. 1999, ApJ, 521, 613
Reid, I. N. & Mahoney, S. 2000, MNRAS, 316, 827
Reid, N. 1992, MNRAS, 257, 257
Reid, N. 1993, MNRAS, 265, 785
Reino, S., de Bruijne, J., Zari, E., d'Antona, F., & Ventura, P. 2018, MNRAS, 477, 3197
Rosenfield, P., Marigo, P., Girardi, L., et al. 2016, ApJ, 822, 73
Röser, S., Demleitner, M., & Schilbach, E. 2010, AJ, 139, 2440
Röser, S., Schilbach, E., Piskunov, A. E., Kharchenko, N. V., & Scholz, R.-D. 2011, A&A, 531, A92
Salaris, M. & Bedin, L. R. 2018, MNRAS, 480, 3170
Schilbach, E. & Röser, S. 2012, A&A, 537, A129
Schneider, A. C., Windsor, J., Cushing, M. C., Kirkpatrick, J. D., & Shkolnik, E. L. 2017, AJ, 153, 196
Skrutskie, M. F., Cutri, R. M., Stiening, R., et al. 2006, AJ, 131, 1163
Smart, R. L., Apai, D., Kirkpatrick, J. D., et al. 2017a, MNRAS, 468, 3764
Smart, R. L., Ioannidis, G., Jones, H. R. A., Bucciarelli, B., & Lattanzi, M. G. 2010, A&A, 514, A84
Smart, R. L., Lattanzi, M. G., Bucciarelli, B., et al. 2003, A&A, 404, 317
Smart, R. L., Marocco, F., Caballero, J. A., et al. 2017b, MNRAS, 469, 401
Soderblom, D. R., Jones, B. F., Stauffer, J. R., & Chaboyer, B. 1995, AJ, 110, 729
Stauffer, J. R., Liebert, J., & Giampapa, M. 1995, AJ, 109, 298
Stauffer, J. R., Liebert, J., Giampapa, M., et al. 1994, AJ, 108, 160
Steele, I. A., Smith, R. J., Rees, P. C., et al. 2004, in Proceedings of SPIE, Vol. 5489, Ground-based Telescopes, ed. J. M. Oschmann, Jr., 679–692
Tabernero, H. M., Montes, D., & González Hernández, J. I. 2012, A&A, 547, A13
Tang, J., Bressan, A., Rosenfield, P., et al. 2014, MNRAS, 445, 4287
Taylor, B. J. 2006a, AJ, 132, 2453
Taylor, M. B. 2006b, in Astronomical Society of the Pacific Conference Series, Vol. 351, Astronomical Data Analysis Software and Systems XV, ed. C. Gabriel, C. Arviset, D. Ponz, & S. Enrique, 666
Terlevich, E. 1987, MNRAS, 224, 193
Tody, D. 1986, in Society of Photo-Optical Instrumentation Engineers (SPIE) Conference Series, Vol. 627, Society of Photo-Optical Instrumentation Engineers (SPIE) Conference Series, ed. D. L. Crawford, 733
Tody, D. 1993, in Astronomical Society of the Pacific Conference Series, Vol. 52, Astronomical Data Analysis Software and Systems II, ed. R. J. Hanisch, R. J. V. Brissenden, & J. Barnes, 173
Tremblay, P.-E., Schilbach, E., Röser, S., et al. 2012, A&A, 547, A99
von Hippel, T. 1998, AJ, 115, 1536
Weidemann, V., Jordan, S., Iben, Jr., I., & Casertano, S. 1992, AJ, 104, 1876
White, R. J., Gabor, J. M., & Hillenbrand, L. A. 2007, AJ, 133, 2524
Wright, E. L., Eisenhardt, P. R. M., Mainzer, A. K., et al. 2010, AJ, 140, 1868
Zuckerman, B. & Song, I. 2004, ARA&A, 42, 685

**Table A.1.** Epochs of observations with IO:I on the Liverpool telescope for Hya02.

| Name | Semester | Epoch |
|------|----------|-------|
|      |          | yyyymmdd |
| Hya02 | CL15B06 | 20150827 |
| Hya02 | CL15B06 | 20150908 |
| Hya02 | CL15B06 | 20151111 |
| Hya02 | CL15B06 | 20160103 |
| Hya02 | CL15B06 | 20160214 |
| Hya02 | CL16B03 | 20160816 |
| Hya02 | CL16B03 | 20160903 |
| Hya02 | CL16B03 | 20160915 |
| Hya02 | CL16B03 | 20161015 |
| Hya02 | CL17B01 | 20171017 |
| Hya02 | CL17B01 | 20180114 |

**Table A.2.** Epochs of observations with IO:I on the Liverpool telescope for Hya10.

| Name | Semester | Epoch |
|------|----------|-------|
|      |          | yyyymmdd |
| Hya10 | CL15B06 | 20150904 |
| Hya10 | CL15B06 | 20150920 |
| Hya10 | CL15B06 | 20151005 |
| Hya10 | CL15B06 | 20151111 |
| Hya10 | CL15B06 | 20160102 |
| Hya10 | CL15B06 | 20160212 |
| Hya10 | CL16B03 | 20160825 |
| Hya10 | CL16B03 | 20160911 |
| Hya10 | CL16B03 | 20160925 |
| Hya10 | CL16B03 | 20161015 |

**Table A.3.** Epochs of observations with IO:I on the Liverpool telescope for Hya09.

| Name | Semester | Epoch |
|------|----------|-------|
|      |          | yyyymmdd |
| Hya19 | CL15B06 | 20150903 |
| Hya19 | CL15B06 | 20150928 |
| Hya19 | CL15B06 | 20151005 |
| Hya19 | CL15B06 | 20151006 |
| Hya19 | CL15B06 | 20151120 |
| Hya19 | CL15B06 | 20160101 |
| Hya19 | CL15B06 | 20160117 |
| Hya19 | CL15B06 | 20160209 |
| Hya19 | CL16B03 | 20160911 |
| Hya19 | CL16B03 | 20160924 |
| Hya19 | CL16B03 | 20161007 |
| Hya19 | CL16B03 | 20161022 |
| Hya19 | CL17B01 | 20171017 |
| Hya19 | CL17B01 | 20180103 |
| Hya19 | CL17B01 | 20180125 |

# Appendix A: Ground-based parallaxes

We list the dates corresponding to the night of observations for the eight targets observed with IO:I on the 2-m robotic Liverpool telescope.

**Table A.4.** Epochs of observations with IO:I on the Liverpool telescope for Hya11.

| Name | Semester | Epoch |
|------|----------|-------|
|      |          | yyyymmdd |
| Hya11 | CL15B06 | 20150906 |
| Hya11 | CL15B06 | 20150920 |
| Hya11 | CL15B06 | 20151114 |
| Hya11 | CL15B06 | 20160103 |
| Hya11 | CL16B03 | 20160826 |
| Hya11 | CL16B03 | 20160911 |
| Hya11 | CL16B03 | 20160925 |
| Hya11 | CL16B03 | 20161015 |
| Hya11 | CL17B01 | 20171019 |
| Hya11 | CL17B01 | 20180119 |

**Table A.5.** Epochs of observations with IO:I on the Liverpool telescope for Hya12.

| Name | Semester | Epoch |
|------|----------|-------|
|      |          | yyyymmdd |
| Hya12 | CL15B06 | 20150906 |
| Hya12 | CL15B06 | 20150927 |
| Hya12 | CL15B06 | 20150928 |
| Hya12 | CL15B06 | 20151012 |
| Hya12 | CL15B06 | 20151120 |
| Hya12 | CL15B06 | 20160102 |
| Hya12 | CL15B06 | 20160118 |
| Hya12 | CL15B06 | 20160210 |
| Hya12 | CL16B03 | 20160817 |
| Hya12 | CL16B03 | 20160905 |
| Hya12 | CL16B03 | 20160918 |
| Hya12 | CL16B03 | 20161020 |
| Hya12 | CL16B03 | 20161022 |
| Hya12 | CL17B01 | 20171011 |
| Hya12 | CL17B01 | 20171018 |
| Hya12 | CL17B01 | 20180104 |
| Hya12 | CL17B01 | 20180123 |

For each target with ground-based parallaxes derived from our Liverpool program, we show two plots. On the left-hand side panels, we plot the residuals in right ascension (bottom) and declination (top), respectively. On the right-hand side panels, we display the relative positions in X and Y with the best fit to derive trigonomatric parallax.

**Table A.6.** Epochs of observations with IO:I on the Liverpool telescope for HyaL5.

| Name | Semester | Epoch |
|------|----------|-------|
|      |          | yyyymmdd |
| HyaL5 | CL15B06 | 20150906 |
| HyaL5 | CL15B06 | 20150921 |
| HyaL5 | CL15B06 | 20151007 |
| HyaL5 | CL15B06 | 20151120 |
| HyaL5 | CL15B06 | 20151218 |
| HyaL5 | CL15B06 | 20160209 |
| HyaL5 | CL16B03 | 20160905 |
| HyaL5 | CL16B03 | 20160923 |
| HyaL5 | CL16B03 | 20161008 |
| HyaL5 | CL16B03 | 20161015 |
| HyaL5 | CL17B01 | 20171018 |
| HyaL5 | CL17B01 | 20180113 |

**Table A.7.** Epochs of observations with IO:I on the Liverpool telescope for CFHT-Hy-20.

| Name | Semester | Epoch |
|------|----------|-------|
|      |          | yyyymmdd |
| CFHT-Hy-20 | CL15B06 | 20150907 |
| CFHT-Hy-20 | CL15B06 | 20150926 |
| CFHT-Hy-20 | CL15B06 | 20151013 |
| CFHT-Hy-20 | CL15B06 | 20151120 |
| CFHT-Hy-20 | CL15B06 | 20160102 |
| CFHT-Hy-20 | CL15B06 | 20160118 |
| CFHT-Hy-20 | CL15B06 | 20160207 |
| CFHT-Hy-20 | CL16B03 | 20160905 |
| CFHT-Hy-20 | CL16B03 | 20160922 |
| CFHT-Hy-20 | CL16B03 | 20161007 |
| CFHT-Hy-20 | CL16B03 | 20161022 |
| CFHT-Hy-20 | CL17B01 | 20171017 |
| CFHT-Hy-20 | CL17B01 | 20180119 |

**Table A.8.** Epochs of observations with IO:I on the Liverpool telescope for CFHT-Hy-21.

| Name | Semester | Epoch |
|------|----------|-------|
|      |          | yyyymmdd |
| CFHT-Hy-21 | CL15B06 | 20150906 |
| CFHT-Hy-21 | CL15B06 | 20150926 |
| CFHT-Hy-21 | CL15B06 | 20151012 |
| CFHT-Hy-21 | CL15B06 | 20151218 |
| CFHT-Hy-21 | CL16B03 | 20160818 |
| CFHT-Hy-21 | CL16B03 | 20160903 |
| CFHT-Hy-21 | CL16B03 | 20160918 |
| CFHT-Hy-21 | CL16B03 | 20161021 |
| CFHT-Hy-21 | CL17B01 | 20171018 |
| CFHT-Hy-21 | CL17B01 | 20180122 |

## Appendix B: The colour-magnitude diagrams

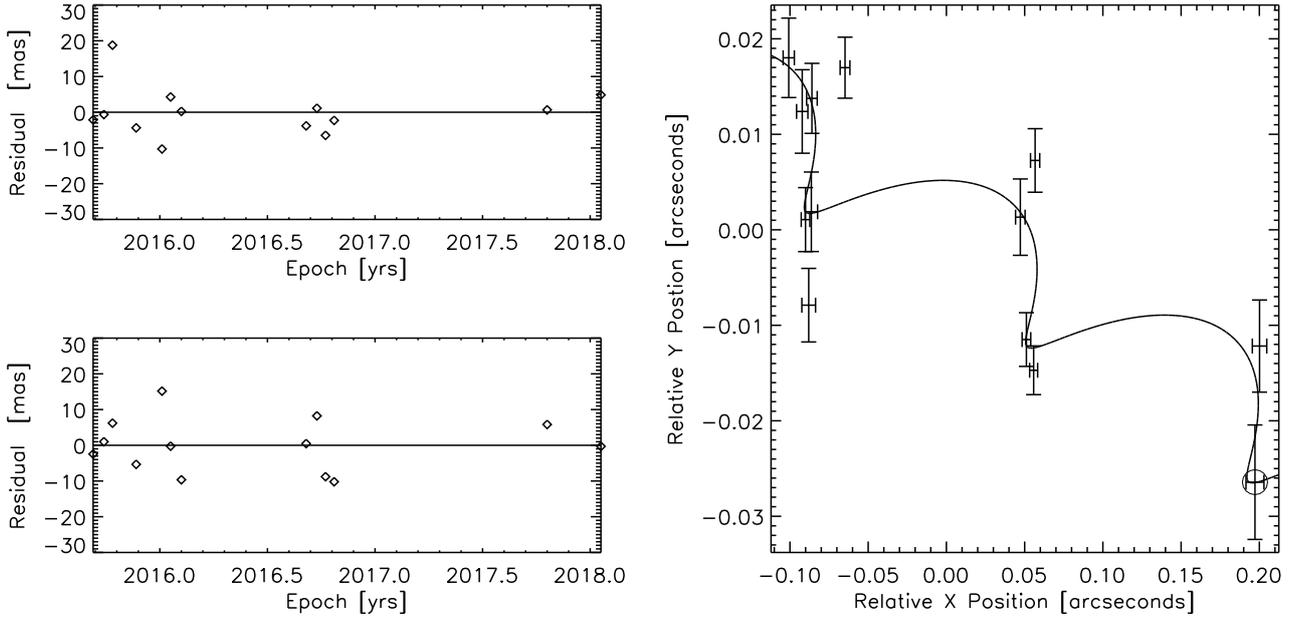

**Fig. A.1.** *Left panels:* Residuals in mas as a function of epoch for right ascension (bottom) and declination (top). *Right panel:* Solution for the parallax determination. We show the results for CFHT-Hy-20, the results for CFHT-Hy-21 are displayed in Fig. 3. The circle dot marks the reference epoch.

## Appendix C: Catalogues

We plan to make public via CDS/Vizier the full table of all candidates within 70° from the cluster center after applying the kinematic analysis described in Sect. 5.1. The full catalogue contains 1764 sources with *Gaia* DR2 data and photometry from several large-scale survey (87, 391, 574, 709 sources within 3.1, 9, 18, and 30 pc from the cluster center, respectively). We expect a much lower membership probability for sources beyond the 30 pc radius because of the size of the cluster. Below we show a subset with some limited *Gaia* properties for space reasons, including source identifier, coordinates, proper motion, parallax, *G* magnitude, galactic coordinates ($b_x$, $b_y$, $b_z$), distance from the center of the cluster, "c" parameter, mass (in M$_\odot$), and radial velocity when available. The full table will include photometry for all candidates from the large-scale surveys discussed in Sect. 2.

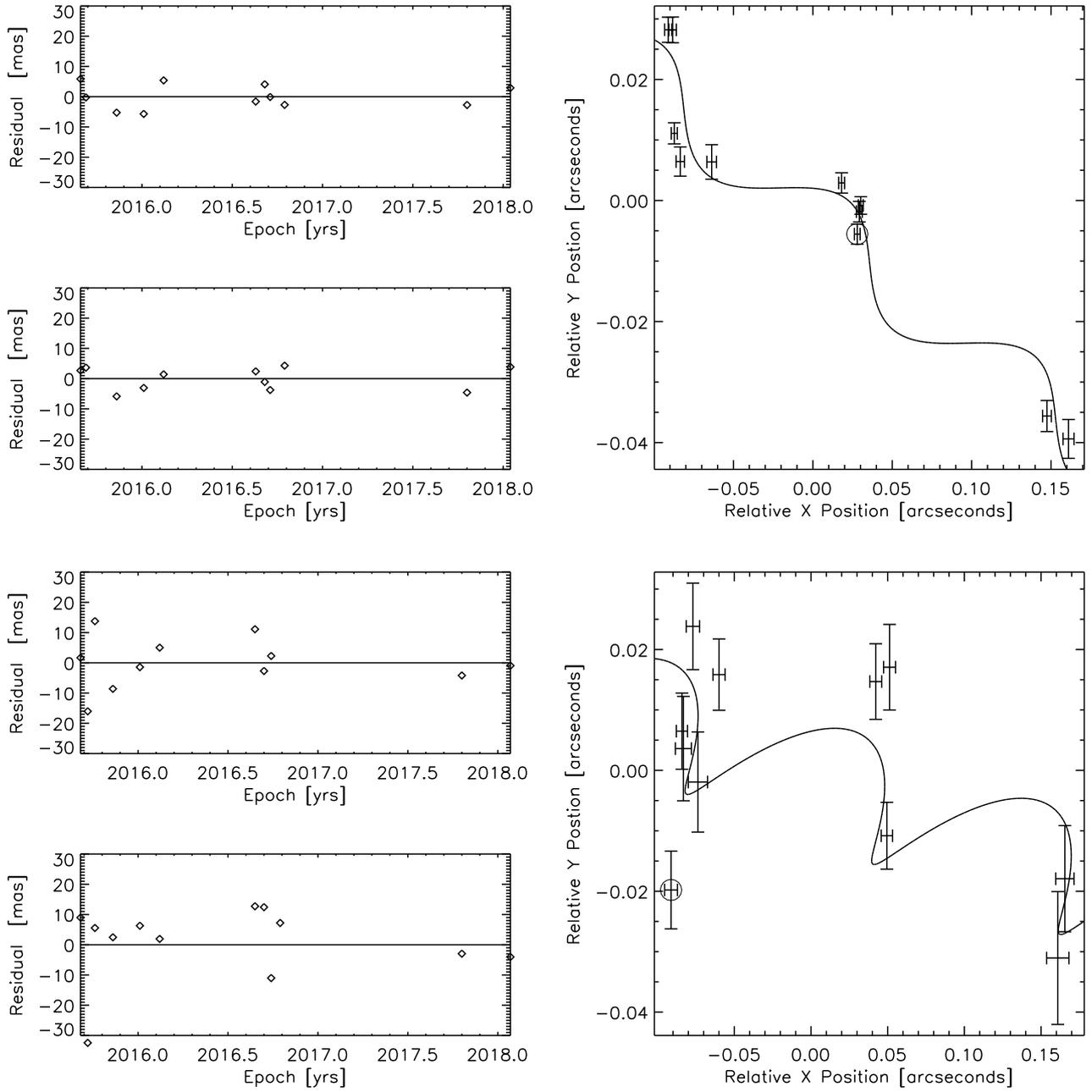

**Fig. A.2.** Same as Fig. A.1 but for Hya02 and Hya10.

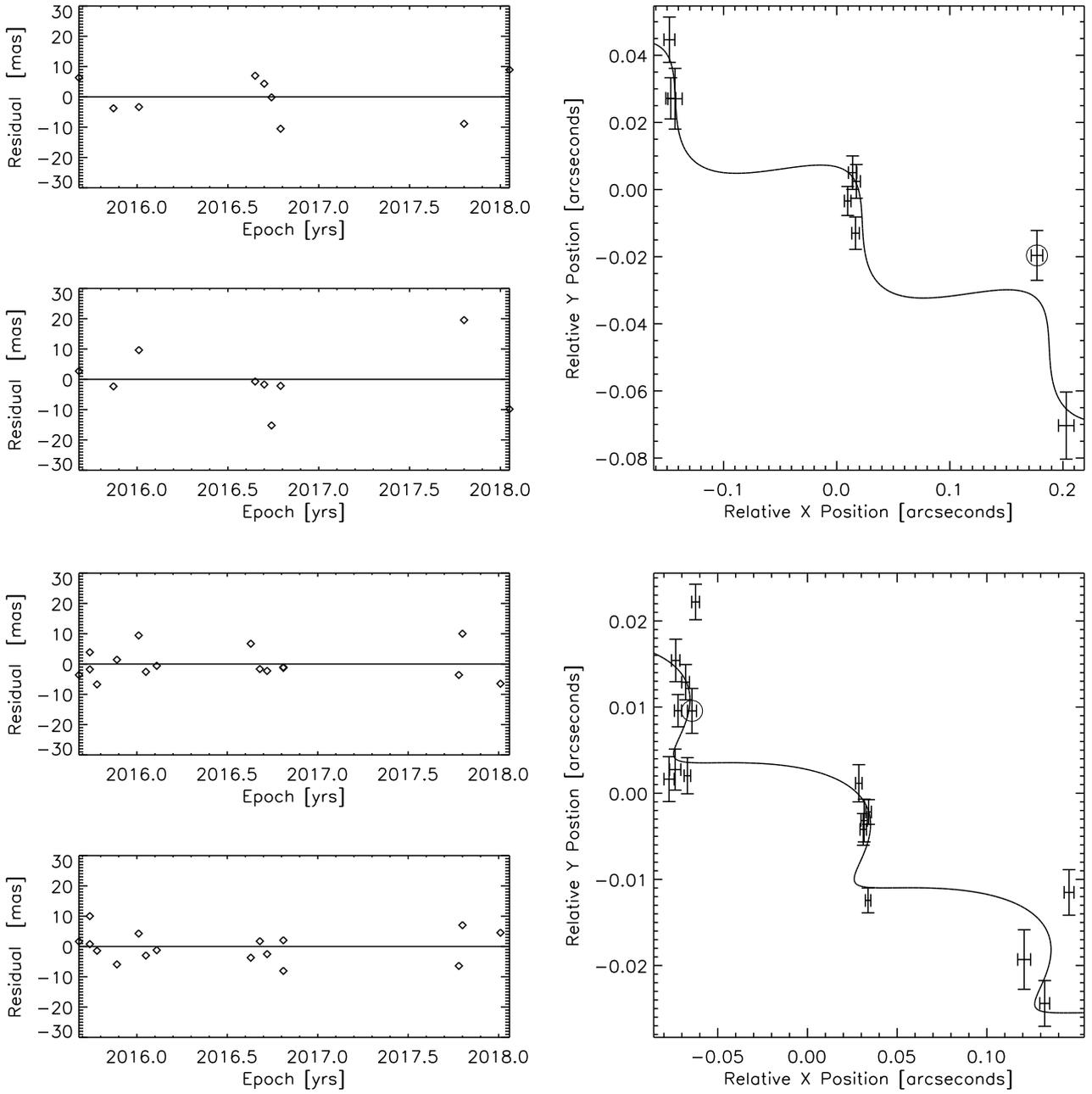

**Fig. A.3.** Same as Fig. A.1 but for Hya11 and Hya12.

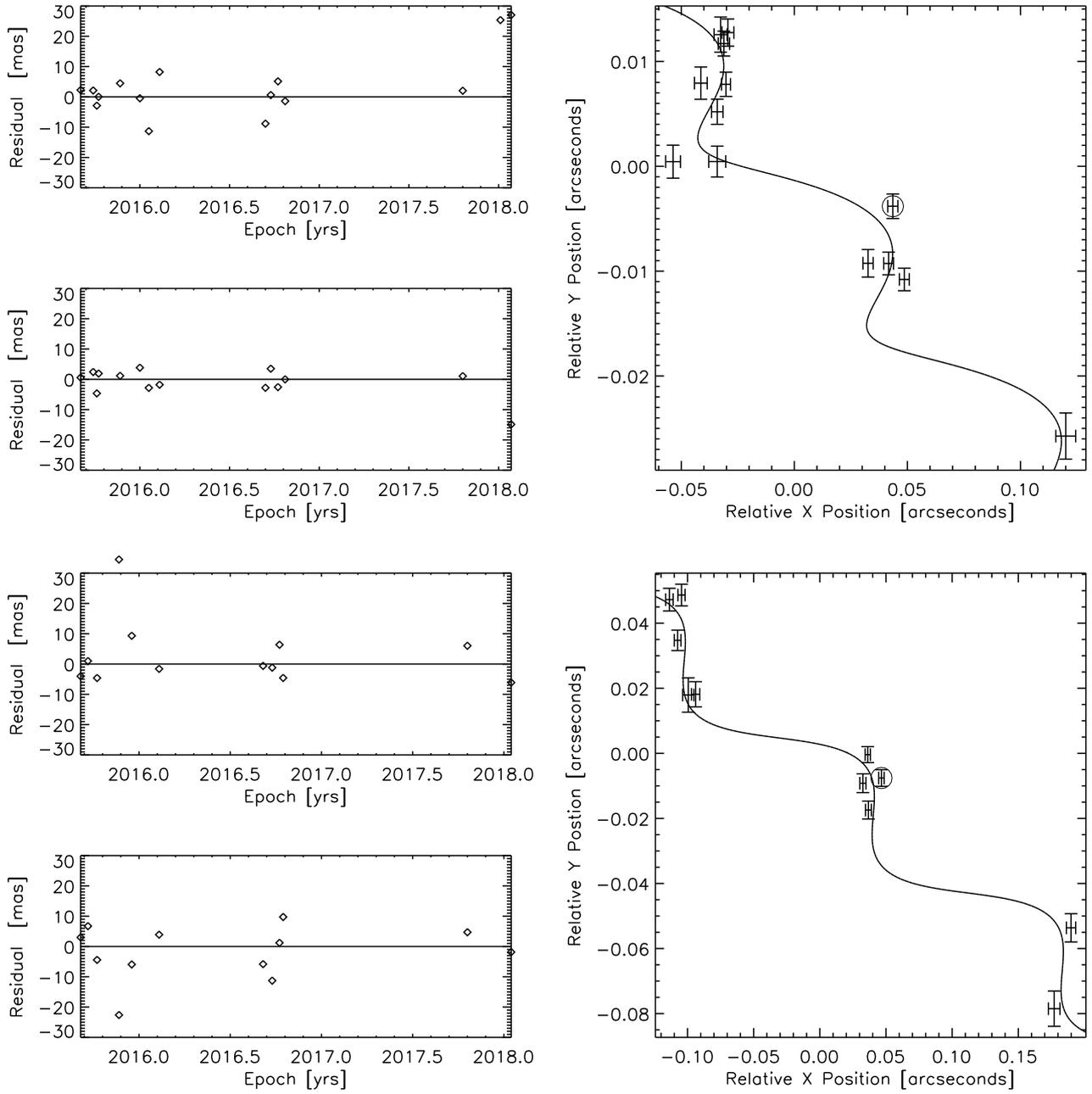

**Fig. A.4.** Same as Fig. A.1 but for Hya19 and HyaL5.

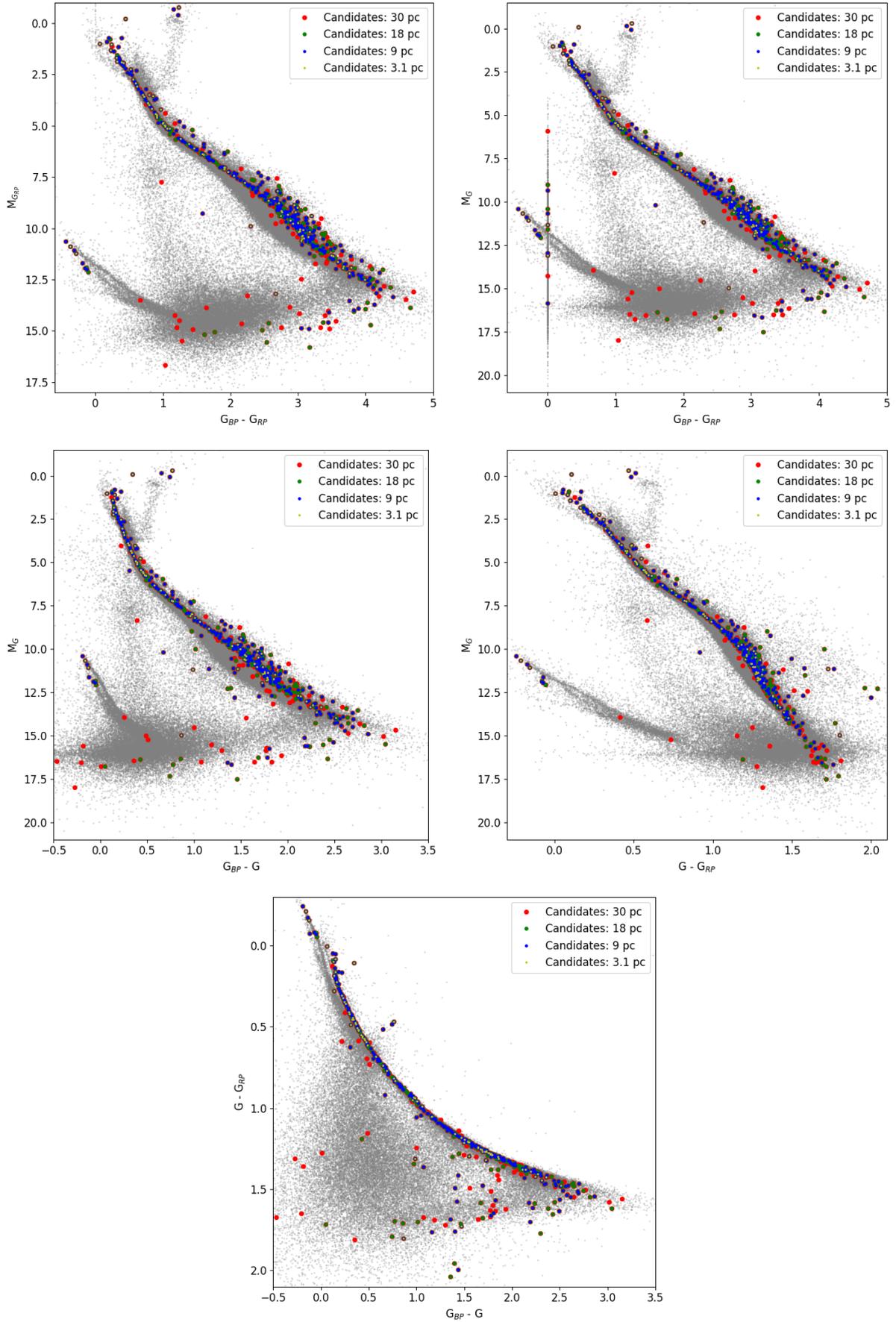

**Fig. B.1.** Colour-magnitude diagrams with *Gaia* photometry only for all candidates within a radius up to 30 pc from the cluster center. All candidates identified by the kinematic method are plotted in black. Members within 3.1, 9, 18, and 30 pc are highlighted in yellow, green, blue, and red, respectively. The bottom panels shows a colour-colour diagram. We added as small grey dots the full *Gaia* catalogue over the 70° radius.

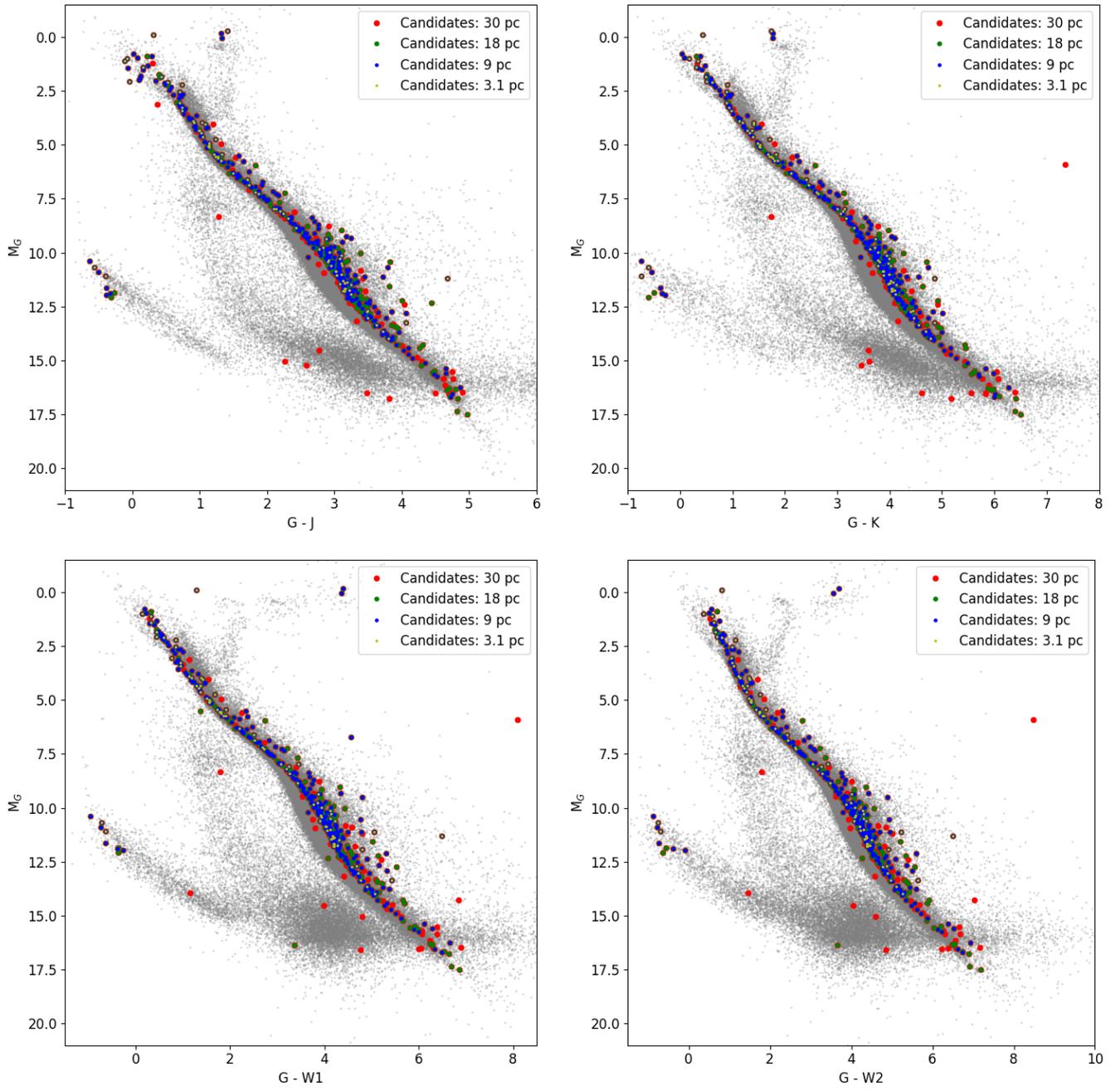

**Fig. B.2.** Colour-magnitude diagrams combining the *Gaia* magnitude with infrared photometry from 2MASS ($J + K_s$) and AllWISE ($W1 + W2$). Symbols are as in Fig. B.1.

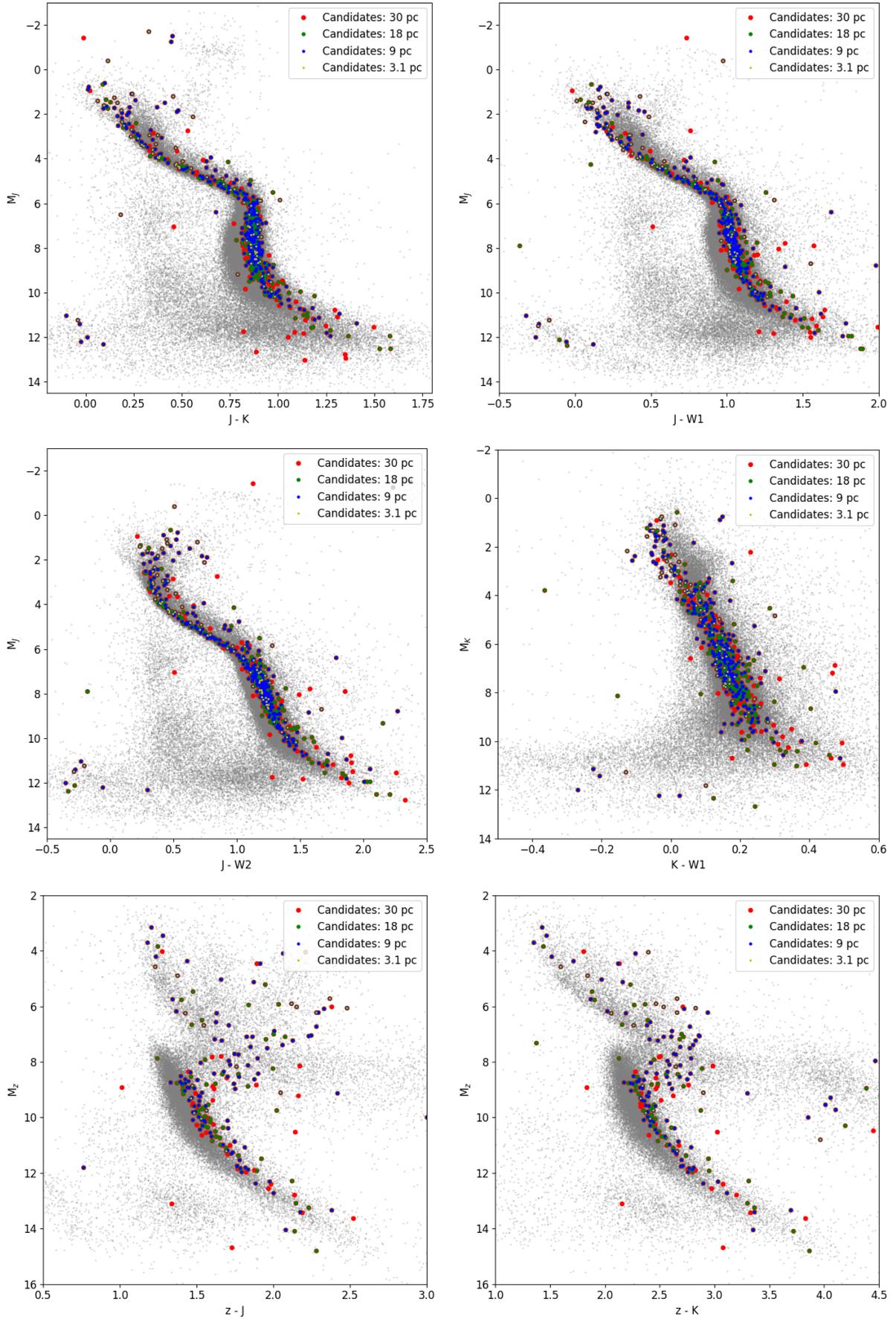

**Fig. B.3.** Colour-magnitude diagrams with non-*Gaia* photometric passbands. Symbols are as in Fig. B.1. Near-infrared filters are from 2MASS, optical ones from Sloan, and mid-infrared data from AllWISE.

**Table C.1.** Catalogue of all 1764 candidates with a velocity consistent with the mean motion of the Hyades and located within a radius of 70° from the cluster center.

| SourceID | RA | DEC | pmRA | pmDEC | Plx | $G$ | $b_x$ | $b_y$ | $b_z$ | d_center | c | Mass | RV |
|---|---|---|---|---|---|---|---|---|---|---|---|---|---|
| | ° | ° | mas/yr | mas/yr | mas | mag | pc | pc | pc | pc | | $M_\odot$ | km s$^{-1}$ |
| 3313714023603261440 | 67.1646712 | 16.9699097 | 102.692 | -26.885 | 20.895 | 14.075 | -44.908 | 0.449 | -17.528 | 0.659 | 0.245 | 0.288 | NaN |
| 3314137846679344128 | 68.0336504 | 17.6643483 | 99.603 | -30.547 | 20.954 | 14.009 | -44.855 | 0.463 | -16.538 | 0.721 | 1.042 | 0.294 | NaN |
| 3314212068010811904 | 67.3769412 | 17.8630271 | 106.261 | -32.129 | 21.275 | 6.780 | -44.128 | 0.907 | -16.589 | 0.959 | 2.159 | 1.329 | 39.679 |
| 3312744219987686400 | 67.1660648 | 15.8707872 | 103.330 | -18.625 | 20.835 | 3.307 | -44.477 | -0.270 | -17.980 | 0.993 | 13.892 | 2.220 | NaN |
| 3312536927686011392 | 67.5363116 | 15.6377391 | 108.920 | -22.137 | 21.120 | 5.591 | -43.988 | -0.602 | -17.664 | 1.091 | 10.509 | 1.634 | NaN |
| 3312904233289409024 | 67.2511010 | 16.3462176 | 104.875 | -27.380 | 21.374 | 13.095 | -43.344 | -0.003 | -17.204 | 1.169 | 2.455 | 0.408 | NaN |
| 3312951748510907392 | 67.4910131 | 16.6727186 | 106.625 | -28.407 | 21.896 | 8.854 | -43.303 | 0.086 | -16.852 | 1.219 | 0.317 | 0.908 | 40.139 |
| 3314109916508903936 | 67.2017334 | 17.2853458 | 108.763 | -28.460 | 21.431 | 7.686 | -43.386 | 0.614 | -16.737 | 1.243 | 5.030 | 1.123 | 39.032 |
| 3312709379213017600 | 66.9000168 | 15.5890888 | 104.992 | -24.069 | 20.925 | 7.274 | -44.141 | -0.317 | -18.182 | 1.247 | 0.119 | 1.221 | 39.395 |
| ... | ... | ... | ... | ... | ... | ... | ... | ... | ... | ... | ... | ... | ... |
| 1997756625600681472 | 353.0966944 | 56.2350598 | 90.382 | 25.004 | 10.001 | 14.985 | -37.529 | 92.613 | -8.670 | 93.079 | 12.183 | 0.376 | NaN |
| 701217290122550912 | 141.6449080 | 32.9007696 | -74.878 | -29.573 | 10.026 | 16.487 | -68.270 | -15.155 | 71.676 | 93.168 | 8.192 | 0.204 | NaN |
| 4772203003708397568 | 81.5885971 | -52.2075581 | 27.343 | 71.587 | 10.253 | 18.582 | -14.902 | -79.934 | -54.702 | 93.317 | 22.080 | 0.108 | NaN |
| 1024580361315032320 | 142.9441526 | 54.7070641 | -140.691 | -122.053 | 10.064 | 20.632 | -67.773 | 22.836 | 70.749 | 93.656 | 15.890 | 0.079 | NaN |
| 396026569162536448 | 2.2201788 | 53.2133036 | 96.766 | 12.982 | 10.472 | 10.001 | -54.187 | 108.670 | -19.507 | 108.951 | 12.377 | 0.983 | 0.414 |
| 4888148944397904896 | 59.2260184 | -29.3694002 | 78.442 | -2.278 | 10.320 | 20.906 | -366.025 | -397.490 | -629.569 | 797.902 | 7.357 | 0.076 | NaN |
| 3117533141727595520 | 97.2884053 | -1.7451372 | 115.532 | -227.258 | 19.145 | 18.357 | 1452.690 | 905.795 | 172.487 | 1760.020 | 10.354 | 0.089 | NaN |
| 2937665168410470912 | 95.5027865 | -21.9249190 | 0.053 | -87.109 | 10.193 | 20.587 | 6395.090 | 7555.060 | 2842.080 | 10330.500 | 0.005 | 0.079 | NaN |
| 109486550395555072 | 44.2464377 | 22.4502375 | 51.672 | -40.243 | 10.053 | 18.489 | 9570.490 | -3972.080 | 6443.110 | 12245.900 | 0.213 | 0.111 | NaN |